\documentclass[colorlinks,allcolors=blue]{jpp}
\usepackage{microtype}
\usepackage{amsmath,array}
\usepackage{algorithm,algpseudocode}
\usepackage[font=it]{subcaption}
\usepackage{orcidlink}
\usepackage{aas_macros}

\allowdisplaybreaks[1]

\def\vb#1{\boldsymbol{#1}}
\def\dd#1{\mathop{}\!\mathrm{d}{#1}}
\def\ol#1{\,\overline{\!{#1}}}
\newcolumntype{C}{>{$}c<{$}}

\def\TPI{Institut f\"ur Theoretische Physik I, Ruhr-Universit\"at Bochum, 44801 Bochum, Germany}
\def\TPIV{Institut f\"ur Theoretische Physik IV, Ruhr-Universit\"at Bochum, 44801 Bochum, Germany}
\def\Oxford{Department of Physics, University of Oxford, Oxford OX1 3PU, UK}

\title{Modelling cosmic-ray transport: magnetised versus unmagnetised motion in astrophysical magnetic turbulence}

\author{%
    Jeremiah Lübke\aff{1,\corresp{\email{jeremiah.luebke@rub.de}}\,\orcidlink{0000-0001-6338-9728}},
    Patrick Reichherzer\aff{2\,\orcidlink{0000-0003-4513-8241}},
    Sophie Aerdker\aff{3\,\orcidlink{0000-0002-4777-4842}},
    Frederic Effenberger\aff{1,3\,\orcidlink{0000-0002-7388-6581}},
    Mike Wilbert\aff{1\,\orcidlink{0000-0003-2831-1583}},
    Horst Fichtner\aff{3\,\orcidlink{0000-0002-9151-5127}}
    \and Rainer Grauer\aff{1\,\orcidlink{0000-0003-0622-071X}}
}

\affiliation{%
    \aff{1}\TPI
    \aff{2}\Oxford
    \aff{3}\TPIV
}

\begin{document}

\maketitle

\begin{abstract}
Cosmic-ray transport in turbulent astrophysical environments remains a multifaceted problem and, despite decades of study, the impact of complex magnetic field geometry --- evident in simulations and observations --- has only recently received more focussed attention. To understand how ensemble-averaged transport behaviour emerges from the intricate interactions between cosmic rays and structured magnetic turbulence, we run test-particle experiments in snapshots of a strongly turbulent magnetohydrodynamics simulation. We characterise particle–turbulence interactions via the gyro radii of particles and their experienced field-line curvatures, which reveals two distinct transport modes: magnetised motion, where particles are tightly bound to strong coherent flux tubes and undergo large-scale mirroring; and unmagnetised motion, characterised by chaotic scattering through weak and highly tangled regions of the magnetic field. We formulate an effective stochastic process for each mode: compound subdiffusion with long mean free paths for magnetised motion, and a Langevin process with short mean free paths for unmagnetised motion. A combined stochastic walker that alternates between these two modes accurately reproduces the mean squared displacements observed in the test-particle data. Our results emphasise the critical role of coherent magnetic structures in comprehensively understanding cosmic-ray transport and lay a foundation for developing a theory of geometry-mediated transport.
\end{abstract}

\section{Introduction}\label{sec:intro}
Turbulence inevitably appears wherever nonlinear flows are at play, which applies to virtually all astrophysical plasmas, such as the solar wind~\citep{Engelbrecht2022}, the interstellar medium~\citep{Elmegreen2004}, or the intracluster medium~\citep{Ryu2012}.
It plays a key role in answering questions about the physics of cosmic rays, such as those regarding transport and energisation processes, as well as the effects of self-generated turbulence and dynamically relevant pressure~\citep{Amato2018}.
Applications, including those reported by \cite{Hopkins2020}, \cite{Dorner2023}, \cite{Ewart2024} and \cite{KamalYoussef2024}, rely on a solid understanding of the transport of fast charged particles through turbulent magnetic fields, which has been intensively studied since the work of \cite{Parker1965}.
However, due to the highly complex nature of plasma turbulence and the strong dependence of cosmic-ray transport on turbulence properties, a comprehensive description is difficult to attain.
This is illustrated by the circumstance that, despite extensive work, existing phenomenological theories are hardly compatible with the available observed cosmic-ray data, as recently discussed by \cite{Kempski2022} and \cite{Hopkins2022a}.

In large parts of the existing literature, the principal paradigm in understanding the interplay between cosmic rays and magnetic turbulence has been gyro-resonance, i.e. scattering of particles by Alfvén waves for comparable gyro radii and wavenumbers~\citep{Kulsrud2005}.
These waves can either be self-generated by streaming cosmic rays~\citep{Skilling1975c}, or emerge as the constituting parts of an extrinsic turbulent cascade.
The latter case is classically treated with quasi-linear theory (QLT), where small-angle scattering in pitch angle caused by gyro-resonance accumulates to a random walk of the particle along a magnetic field line~\citep{Jokipii1966,Mertsch2020,Reichherzer2020,Els2024a}.
However, this paradigm views magnetic turbulence merely as an ensemble of linear space-filling magnetohydrodynamic (MHD) waves with a given energy spectrum.
This is incompatible with strong MHD turbulence~\citep{Meyrand2016}, which tends to form \textit{coherent structures}, i.e. ordered patches characterised by strong alignment and reduced nonlinearity~\citep[see, e.g.][]{Perez2009,Matthaeus2015}.
A geometric perspective on magnetic turbulence is also advocated by, e.g. \cite{Grauer1994}, \cite{Politano1995b}, \cite{Grauer2000}, \cite{minini-pouquet-etal:2006}, \cite{Boldyrev2006} and \cite{Malara2021}.
Coherent structures have been extensively observed in the solar wind~\citep{tu-marsch:1995,Khabarova2021,Vinogradov2024},
and numerically studied in the context of cosmic-ray acceleration~\citep{Arzner2006,Lemoine2021,Pezzi2022,Pugliese2023}.

While strong turbulence as a mixture of coherent and chaotic structures marginally reproduces the expected energy spectrum, such an averaging argument does not readily apply to cosmic-ray transport, because the transport coefficients depend strongly on the geometry and distribution of these structures.
This was demonstrated by \cite{Shukurov2017}, who found longer mean free paths (MFPs) of test particles in a fluctuating dynamo compared with random-phase synthetic turbulence with the same energy spectrum.
Additionally, coherent structures may act as non-resonant magnetic mirrors~\citep{Chandran1999,Albright2001,Bell2025} and govern field-line wandering, which are two important mechanisms observed in test-particle experiments in snapshots of MHD simulations~\citep{Beresnyak2011,Xu2013,Cohet2016,Zhang2024}.
The wandering of field lines is also important for the effective diffusion of streaming cosmic rays~\citep{Sampson2023}.
Further, sharply curved magnetic field lines, which may arise at the interfaces between coherent structures, were recently invoked as an effective scattering agent for cross-field transport~\citep{Kempski2023,Lemoine2023}.
Even though a micro-physical theory of cosmic-ray transport beyond gyro-resonance is not yet available, the general understanding of magnetic turbulence has improved significantly in recent years (see \citealt{Schekochihin2022} for a comprehensive review), thus providing the foundations for the development of such a theory.
That is not to say that gyro-resonance is entirely obsolete; rather, actual transport behaviour likely results from the interplay of several different mechanisms with varying contributions, depending on the turbulent properties of a given astrophysical system, which may vary in space and scale.

In this paper, we explore how a theory of cosmic-ray transport mediated by turbulence-induced geometry may look like.
We do this by means of test-particle experiments in an MHD simulation of a fluctuating dynamo, where a dynamically dominant flow amplifies an initially small magnetic field and produces pronounced coherent flux tubes~\citep{Schekochihin2004a,Rincon2019,Seta2020}.
This process is believed to play an important role in the generation of magnetic fields in galaxies~\citep{Rieder2017,Gent2024} and the intracluster medium~\citep{Vazza2018,Steinwandel2024}, with initial seed fields possibly provided by the Weibel instability~\citep{Sironi2023,Zhou2024}.
To understand the connection between test-particle motion and magnetic field geometry, we investigate magnetic moment variations and mean squared displacements (MSDs) conditional on the gyro radii of the particles and their experienced field-line curvature.
This reveals two distinct modes of transport, namely \textit{magnetised motion}, where particles are closely bound to a strong and ordered magnetic field inside coherent structures, and \textit{unmagnetised motion} consisting of chaotic scattering through a weak and tangled magnetic field.
This description can be made more precise by fitting a stochastic model to the test-particle motion, where field-line wandering and mirroring in the magnetised case are represented by compound subdiffusion~\citep{Balescu1994,Qin2002a,Neuer2006,Minnie2009}, and the chaotic scattering in the unmagnetised case is represented by a three-dimensional (3-D) Langevin equation~\citep{Chandrasekhar1943,Bian2024}.
We conclude our treatment by discussing implications and open questions towards a proper theory for cosmic-ray transport.

The paper proceeds by presenting the MHD simulation and test-particle experiments in Section~\ref{sec:simu}, followed by the stochastic model and fitting procedure in Section~\ref{sec:model}.
We then discuss the implications of our results in Section~\ref{sec:discussion}, and conclude with an outlook in Section~\ref{sec:outlook} and a summary in Section~\ref{sec:summary}.

\section{MHD and test-particle simulations}\label{sec:simu}
\subsection{MHD simulation set-up}\label{sec:simusetup}
We perform a simulation of incompressible visco-resistive MHD turbulence in a 3-D periodic box.
The governing equations of the flow field~$\vb{u}$ and the magnetic field~$\vb{B}$, which are given by~\citep{Biskamp2003}
\begin{subequations}\label{eq:mhd}
    \begin{gather}
        \frac{\partial\vb{u}}{\partial t} +\vb{u} \bcdot \nabla \vb{u} = \vb{B} \bcdot \nabla \vb{B}
        -\nabla p -\nu_h(-\Delta)^h \vb{u} +\vb{f}, \quad \nabla\bcdot\vb{u} = 0, \\
        \frac{\partial\vb{B}}{\partial t}+\vb{u}\bcdot\nabla\vb{B} -\vb{B}\bcdot\nabla\vb{u} =
        -\eta_h(-\Delta)^h \vb{B}, \quad \nabla\bcdot\vb{B} = 0,
    \end{gather}
\end{subequations}
with hyper-viscosity~$\nu_h$ and hyper-resistivity~$\eta_h$ of order~$h$,
are solved with our pseudo-spectral code \textit{SpecDyn}~\citep{Wilbert2022,WilbertPhd2023}.
The equations are solved on a 3-D uniform grid with~$1024^3$ points and periodic boundary conditions. The length of the domain is~$L_\mathrm{box}=2\pi$ and the time scale of the flow is~$T_\mathrm{eddy}=L_\mathrm{box}/u_\mathrm{rms}$. We set~$\nu_h=\eta_h$, resulting in the magnetic Prandtl number~$Pr_m=1$.
The magnetic field~$\vb{B}$, which is specified in Alfvénic units~$[B]=[u]$, is initialised with a small-amplitude seed field with zero magnetic helicity and zero net magnetic flux.
The flow field~$\vb{u}$ is driven by a large-scale force density~$\vb{f}$, operating on the wavenumber band~$1\le k\le2$, with random amplitudes, which are~$\delta$-correlated in time to ensure constant power injection. The force density is constructed such that the injected net cross-helicity is zero~\citep{Alvelius1999}.
Due to the chosen wavenumber band of the forcing, the box contains only one flow correlation cell.
We run the simulations until the magnetic energy has been amplified to a statistically saturated state, and then record eight snapshots of~$\vb{u}(\vb{x})$ and~$\vb{B}(\vb{x})$ separated in time by~$T_\mathrm{eddy}$.

We simulate equations~(\ref{eq:mhd}) with~$h=1$ as the baseline case and with~$h=2$ as the production case.
The parameters of the simulations are listed in table~\ref{tab:mhdparams}.
Setting~$h=2$ results in a sharper dissipative cut-off and, thus, an extended inertial range, which is reflected by the Reynolds numbers listed in table~\ref{tab:mhdparams}. Differences in the spectra and geometry are further reflected by the characteristic scales listed in table~\ref{tab:scales}.
For~$h=2$ the magnetic energy saturates at a higher level compared with~$h=1$~\citep[see also][]{Brandenburg2002}; however, this does not matter for our test particle simulations, because we normalise the magnetic field to unit root-mean-square (r.m.s.) strength.
Despite differences in the flux rope geometries, as indicated in table~\ref{tab:scales}, we do not observe significant differences in our conditional particle statistics.
Thus, subsequently, we only show the results for the case~$h=2$, which additionally allows for the meaningful inclusion of lower-energy particles due to the shorter dissipation scale.

See also Section~\ref{sec:simuresults} for additional discussion of the geometric length scales.

\begin{table}
    \centering
    \begin{tabular}{C@{\hskip 9pt}*{5}{C}}
h & w & \nu_h,\eta_h & \nu_\mathrm{eff},\eta_\mathrm{eff} & w_\mathrm{rms} & Re_{w,T} \\[3pt]
1 & u & 1.2\times10^{-3} & 1.2\times10^{-3} & 2.423 & 433.463 \\
1 & B & 1.2\times10^{-3} & 1.2\times10^{-3} & 1.415 & 94.379 \\
2 & u & 3.6\times10^{-8} & 6.2\times10^{-4} & 2.868 & 493.659 \\
2 & B & 3.6\times10^{-8} & 5.4\times10^{-4} & 2.079 & 217.751
    \end{tabular}
    \caption{MHD simulation parameters, for grid size~$1024^3$, box length~$L_\mathrm{box}=2\upi$, and fields~$w\in\{u,B\}$.
    The maximal resolved de-aliased wavenumber is~$k_\mathrm{max}=341$.
    The Reynolds numbers are based on the Taylor scale~$k_{w,T}$ reported in table~\ref{tab:scales}, and for~$h=2$, on the effective viscosity~$\nu_\mathrm{eff}$ and effective resistivity~$\eta_\mathrm{eff}$~\citep{Haugen2004b}.}
    \label{tab:mhdparams}
\end{table}

\begin{table}
    \centering
    \begin{tabular}{C@{\hskip 9pt}*{7}{C}}
        h & w & k_{w,\mathrm{corr}} & k_{w,T} & k_{w,\mathrm{diss}} & k_\parallel & k_{\vb{B}\times\vb{j}} & k_{\vb{B}\cdot\vb{j}} \\[3pt]
        1 & u & 1.868 & 4.658 & 190.86 & - & - & - \\
        1 & B & 7.811 & 12.492 & 238.822 & 1.777 & 15.04 & 11.769 \\
        2 & u & 2.139 & 9.37 & 409.148 & - & - & - \\
        2 & B & 8.983 & 17.682 & 511.787 & 3.233 & 20.499 & 18.142
    \end{tabular}
    \caption{Characteristic length scales of the simulations. We report the correlation scale~$k_{w,\mathrm{corr}}$, Taylor scale~$k_{w,T}$ and Kolmogorov dissipation scale~$k_{w,\mathrm{diss}}$ for both fields~$w\in\{u,B\}$. Additionally, we show for the magnetic field, the characteristic parallel scale~$k_\parallel$, reversal scale~$k_{\vb{B}\times\vb{j}}$ and perpendicular scale~$k_{\vb{B}\cdot\vb{j}}$ \citep[see][]{Schekochihin2004a}.}
    \label{tab:scales}
\end{table}

\begin{table}
    \def\swB{2}
    \def\swL{1}
    \def\mcB{4}
    \def\mcL{3}
    \def\ismB{5}
    \def\ismL{6}
    \def\icmB{7}
    \def\icmL{\ismL}
    \centering
    \begin{tabular}{lccCC}
        System & $B_\mathrm{rms}$ & $L_\mathrm{corr}$ & r_g & E \\[3pt]
        solar wind &  5 nT$^{(\swB)}$ & 0.01 au$^{(\swL)}$ & (24.5,\cdots,6.14)\times10^{-5}\ \mathrm{au} & (0.645,\cdots,0.041)\ \mathrm{MeV} \\
        molecular clouds & 1 nT$^{(\mcB)}$ & 16 mpc$^{(\mcL)}$ & (0.393,\cdots,0.0982)\ \mathrm{mpc} & (2.31,\cdots,0.577)\ \mathrm{TeV} \\
        interstellar medium & 0.1 nT$^{(\ismB)}$ & 100 pc$^{(\ismL)}$ & (2.45,\cdots,0.614)\ \mathrm{pc} & (1.45,\cdots,0.361)\ \mathrm{PeV} \\
        intracluster medium & 0.5 nT$^{(\icmB)}$ & 10 kpc$^{(\icmL)}$ & (0.245,\cdots,0.0614)\ \mathrm{kpc} & (0.723,\cdots,0.181)\ \mathrm{EeV} \\
    \end{tabular}
    \caption{Cosmic-ray protons affected by our considerations with~$\hat{\omega}_g=64,\cdots,256$, characterised by the r.m.s.~field strength~$B_\mathrm{rms}$ and turbulence correlation scale~$L_\mathrm{corr}\sim L_\mathrm{box}$ with values taken from: 
    (\swL) \citet{Weygand-etal-2011};
    (\swB) \citet{Weygand-etal-2013};
    (\mcL) \citet{Houde-etal-2009};
    (\mcB) \citet{Crutcher-2012};
    (\ismB) \citet{Jansson-Farrar-2012};
    (\ismL) \citet{Seta2020}; and
    (\icmB) \citet{DominguezFernandez-2020}.
    Note that values found in the literature tend to be widely spread due to the heterogeneous nature of astrophysical environments, as well as due to intrinsic observational uncertainties.
    The listed values serve merely as order-of-magnitude estimates.
    Also note that our fluctuating dynamo turbulence is not applicable to the anisotropic solar wind~\citep{Chen2020b,Wang2024}, and that the reported particle energies are non-relativistic with~$V_0=(0.037,\cdots,0.009)\,c$.
    For these reasons, the solar wind values are shown only for reference.}
    \label{tab:physics}
\end{table}

\subsection{Test particle set-up}\label{sec:simusetup2}
We then study the motion of cosmic rays in these static snapshots by integrating the Newton-Lorentz equations (see Appendix~\ref{app:lorentz})
\begin{equation}
    \dot{\vb{X}}_t=\vb{V}_t, \quad \dot{\vb{V}}_t=\hat{\omega}_g\vb{V}_t\times\vb{B}(\vb{X}_t)
    \label{eq:lorentz}
\end{equation}
with the volume-preserving Boris scheme~\citep{boris:1971,Ripperda2018}.
This test-particle approach is justified by assuming that relativistic cosmic rays with~$V\approx c$ move in front of a non-relativistic plasma background with~$u_\mathrm{rms}\ll c$, such that the time scales of the MHD and test particle simulations are well separated, i.e.~$T_\mathrm{particle}\ll T_\mathrm{eddy}$ with $T_\mathrm{particle}=L_\mathrm{box}/V_0$.
We can then neglect the influence of the electric field, implying conservation of energy for our test particles.
The particles are parametrised by the normalised r.m.s.~gyro frequency
\begin{equation}
    \hat{\omega}_g=\frac{qB_\mathrm{rms}}{\gamma m}\frac{L_\mathrm{box}}{V_0},
\end{equation}
which includes the amplitudes~$V_0$ of the particle velocity and~$B_\mathrm{rms}$ of the magnetic field; equation~(\ref{eq:lorentz}) is accordingly normalised to~$\|\vb{V}\|=1$ and~$B_\mathrm{rms}=\langle\|\vb{B}\|^2\rangle^{1/2}=1$.
The parameter~$\hat{\omega}_g$ denotes how strongly particles are coupled to the magnetic field and inversely encodes their energy via~$E=\gamma mc^2=qB_\mathrm{rms}L_\mathrm{box}c/\hat{\omega}_g$, with~$V_0=c$.
Table~\ref{tab:physics} provides a mapping to specific astrophysical systems, based on typical values of the magnetic field strengths and turbulence correlation scales.
Due to limited numerical resolution and the requirement to constrain the particle gyro radii to the range of resolved turbulent fluctuations, the representable particle energies are very high for the respective physical systems. How our findings extend to lower energies, such as galactic GeV protons, can only be speculated about, as we attempt in Section~\ref{sec:discussion-asymptotic}.

\begin{figure}
    \centering
    \includegraphics[width=0.5\linewidth]{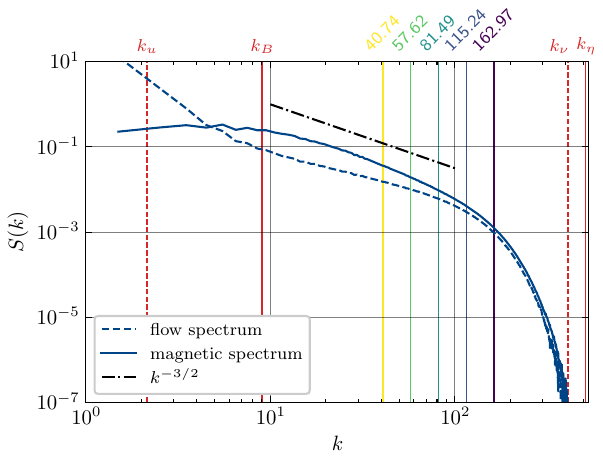}
    \caption{Radially averaged power spectra of flow and magnetic field in the statistically saturated state of the simulation with~$h=2$.
    Indicated are the wavenumbers of the integral scales~$k_u=2.139$ and~$k_B=8.983$, the effective Kolmogorov dissipation scales~$k_\nu=409.148$ and~$k_\eta=511.787$, as well as the r.m.s.~gyro wavenumbers~${\left(\upi/2\right)^{-1}}{\hat{\omega}_g}$ of the considered gyro frequencies~$\hat{\omega}_g=64; 90.51; 128; 181.019; 256$.}
    \label{fig:mhdspectrum}
\end{figure}

The typical gyro period is given by~$T_g=2\upi\,\hat{\omega}_g^{-1}$ and we employ a fixed time step~$\Delta t=10^{-2}\,T_g$ for the integration of equation~(\ref{eq:lorentz}).
The r.m.s. and instantaneous normalised gyro radii are respectively estimated by~$\langle\hat{r}_g\rangle=\frac{\upi}{2}\hat{\omega}_g^{-1}$ and~$\hat{r}_{g,t}=\sqrt{1-\mu_t^2}\,\hat{\omega}_g^{-1}B^{-1}(\vb{X}_t)$, where~$\mu_t=\hat{\vb{V}}_t\bcdot\hat{\vb{B}\,}\!(\vb{X}_t)$ denotes the pitch angle cosine.
We select five values between~$\hat{\omega}_g=256$ and~$\hat{\omega}_g=64$, where particles with larger values follow field lines more closely, whereas particles with smaller values average magnetic structures more coarsely and exhibit behaviour akin to a random walk.
The respective gyro scales~$\langle\hat{r}_g\rangle^{-1}$ are shown in figure~\ref{fig:mhdspectrum} in relation to the radially averaged energy spectra of the flow and magnetic field.

We solve equation~(\ref{eq:lorentz}) with our \textit{ParticlePusher} code, which is able to record binned statistics of arbitrary quantities along particle trajectories.
We use the local gyro average of a quantity~$Q$ along a particle trajectory~$\vb{X}_t$, which is defined as
\begin{equation}
    \bar{Q}(\vb{X}_t)=\frac{1}{\tilde{T}_g(\vb{X}_t)}\int\limits_0^{\tilde{T}_g(\vb{X}_t)}Q(\vb{X}_{t-t'})\dd{t'},
    \label{eq:gyroavg}
\end{equation}
where the local gyro period is estimated as~$\tilde{T}_g(\vb{X}_t)=2\upi\left(\frac{1}{T_g}\int_0^{T_g}\hat{\omega}_gB(\vb{X}_{t-t'})\dd{t'}\right)^{-1}$.
In the magnetised limit~$r_g\ll B/\nabla B$, particles coherently gyrate along an ordered field and~$\bar{Q}$ describes the well-defined gyro-centre motion,
while in the unmagnetised limit~$r_g\gg B/\nabla B$, particles exhibit highly chaotic motion and~$\bar{Q}$ quantifies the competition between the particle's inertia and the quickly varying Lorentz force.
Analogously, we define the local gyro variance as
\begin{equation}
    \ol{\delta Q}^2(\vb{X}_t)=\ol{\left(Q(\vb{X}_t)-\bar{Q}(\vb{X}_t)\right)^2}.
\end{equation}

To investigate scattering and transport behaviour of particles in the magnetised and unmagnetised regimes, we consult the geometric picture of \cite{Lemoine2023} and \cite{Kempski2023} based on the field-line curvature
\begin{equation}
    \kappa=\frac{\|\vb{B}/B\times(\vb{B}\bcdot\nabla\vb{B})\|}{B^2},
    \label{eq:curv}
\end{equation}
where~$\kappa r_g\ll1$ corresponds to the magnetised case,~$\kappa r_g\gg1$ corresponds to the unmagnetised case and for~$\kappa r_g\sim1$ order-unity variations of the magnetic moment,
\begin{equation}
    M=\frac{\gamma m V_\perp^2}{2B}=\frac{\gamma m \left(1-\mu^2\right)}{2B}
\end{equation}
are expected.
Based on these definitions, we record the average of the relative variation of the magnetic moment conditional on the field-line curvature and particle gyro radius, \begin{equation}
    \left\langle \frac{\ol{\delta M}}{\ol{M}} \middle| \bar{\kappa}, \bar{r}_g \right\rangle.
    \label{eq:dM}
\end{equation}
Further, we record spatial mean squared displacements (MSDs) of particles
\begin{equation}
    \langle\Delta X^2_\tau\,|\,\mathrm{cond.}\rangle=\langle\|\vb{X}_{t+\tau}-\vb{X}_t\|^2\,|\,\mathrm{cond.}\rangle,
    \label{eq:dX}
\end{equation}
conditional on the magnetisation criteria~$\bar{\kappa}\bar{r}_g<1$ and~$\bar{\kappa}\bar{r}_g>1$, as well as an unconditional baseline.
We assume stationarity~$\langle\|\vb{X}_\tau-\vb{X}_0\|^2\rangle=\langle\|\vb{X}_{t+\tau}-\vb{X}_t\|^2\rangle$, expressed by the local time scale~$\tau$.
The MSD, obtained from the displacement of individual particles, measures the time-dependent spread of the underlying particle distribution function,
and classifies the motion as super-, normal or sub-diffusive, depending on~$\langle\Delta X^2_\tau\rangle\sim\tau^\alpha$ scaling with~$\alpha>1$,~$=1$ or~$<1$.
For~$\alpha=1$, the classical diffusion coefficient is given by~$D_\infty=\lim_{\tau\to\infty}\langle\Delta X^2_\tau\rangle/2\tau$~\citep{Metzler2000}.

The unconditional MSD should be treated with care, because it averages over many different length scales, thus hiding relevant physical processes, which we address by conditioning on the magnetisation criterion.
Further, since our simulation box only contains one flow correlation cell, and the parallel scale of the magnetic field~$k_\parallel$ is comparable to the box size,~$D_\infty$ converges only after particles traverse the periodic domain multiple times. In doing so, particles likely re-enter the box at a different position and thus sample different regions of the magnetic field, which results in a meaningful albeit biased value for~$D_\infty$.

To obtain the conditional statistics, we simulated for each~$\hat{\omega}_g$, in each of the eight statistically independent MHD snapshots, $400\,000$ independent test-particle trajectories for $1000$ gyro periods.
The unconditional baseline MSD is, per snapshot, based on $40\,000$ independent test-particle trajectories with lengths of $10\,000$ gyro periods.
This ensures that the tails of the joint density~$p(\bar{\kappa},\bar{r}_g)$ are well resolved.
However, the conditional MSD at large time scales is hard to resolve properly, even with increased sample sizes.
This matter is discussed in more detail in Section~\ref{sec:model-fit}.

\subsection{MHD simulation results}\label{sec:simuresults}

\begin{figure}
    \centering
    \includegraphics[width=\linewidth]{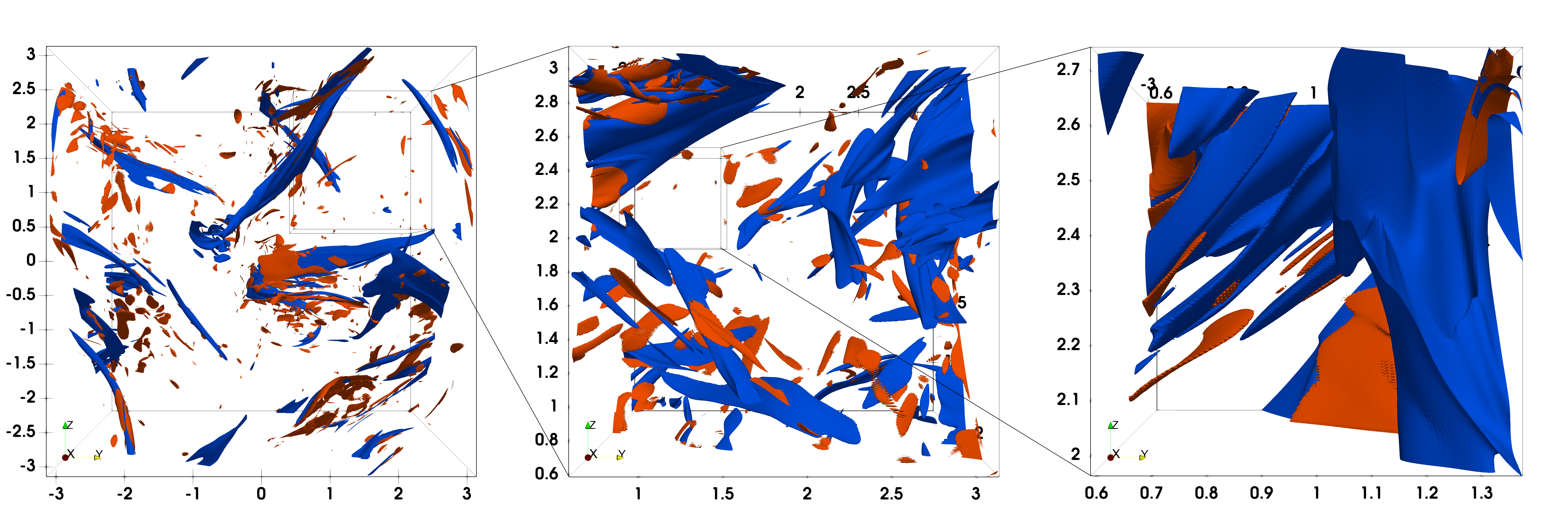}
    \caption{Isosurfaces of the magnetic field strength~$B$ (\textit{\color{blue} blue}) and the current density magnitude~$j=\|\nabla\times\vb{B}\|$ (\textit{\color{red} red}).
    \textit{(left)}~Whole box with~$B_\mathrm{iso}/B_\mathrm{max}=0.7$ and~$j_\mathrm{iso}/j_\mathrm{max}=0.418$.
    \textit{(middle)}~Cutout with~$B_\mathrm{iso}/B_\mathrm{max}=0.489$ and~$j_\mathrm{iso}/j_\mathrm{max}=0.303$.
    \textit{(right)}~Cutout with~$B_\mathrm{iso}/B_\mathrm{max}=0.245$ and~$j_\mathrm{iso}/j_\mathrm{max}=0.115$.
    The subscript \textit{iso} denotes the value at which the isosurfaces are drawn.
    The structures of the magnetic isosurfaces correspond to flux tubes, as indicated in figure~\ref{fig:trajectories}, which are amplified by the fluctuating dynamo action.
    Most of the magnetic energy is concentrated on large scales in a few intense flux tubes, while small scales reveal less intense and tightly folded flux tubes.
    Current sheets appear in close proximity to intense flux tubes and are embedded between folds.}
    \label{fig:iso}
\end{figure}

We start by inspecting the magnetic field snapshots extracted from the statistically saturated phase of the MHD simulation.
Figure~\ref{fig:iso} shows isosurfaces of the magnetic field strength~$B$ and the current density magnitude~$j=\|\nabla\times\vb{B}\|$ at different zoom levels in the simulation box. The isosurfaces of~$B$ reveal characteristic structures of the saturated fluctuating dynamo~\citep{MillerPhd2019,Seta2020}, which can be described as long flattened tubes with length~$l$, width~$w$ and height~$h$, ordered as~$l \gg w>h$.
As indicated in figure~\ref{fig:trajectories}, they contain coherent bundles of field lines and approximately represent flux surfaces.
Thus, henceforth, we refer to these coherent structures as \textit{flux tubes}.
The connection between field strength~$B$ and coherent geometry with small curvatures~$\kappa$ is formally expressed by the anti-correlation between the two quantities~$B\sim\kappa^{-1/2}$~\citep{Schekochihin2004a}.

\begin{figure}
    \centering
    \includegraphics[width=.85\linewidth]{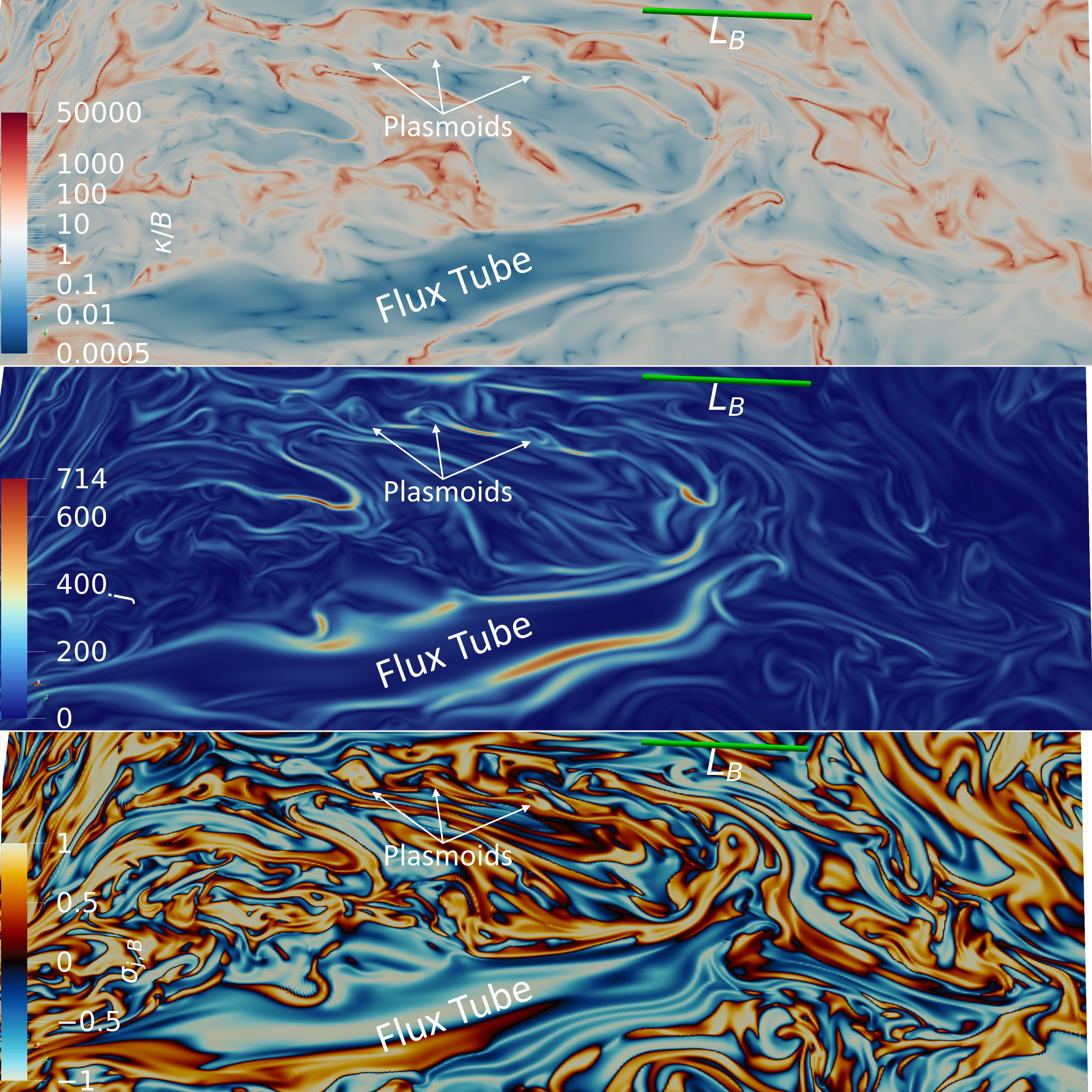}
    \caption{Slices through a magnetic flux tube, surrounded by tight folds, current sheets and plasmoids.
    \textit{(top)}~Field-line curvature divided by the field strength~$\kappa/B$ for comparison with the magnetisation criterion~$\kappa r_g\sim 1$ with~$r_g\propto B^{-1}$.
    \textit{(centre)}~Magnitude of the current density~$j$ indicating intense current sheets.
    \textit{(bottom)}~Alignment between the magnetic field and current density~$\sigma_{j,B}=\hat{\vb{j}\,}\!\bcdot\hat{\vb{B}\,}\!$ indicating 
    cellularisation
    into approximately force-free patches.
    Further indicated are the correlation scale of the magnetic field and the locations of the flux tube and example plasmoids.}
    \label{fig:slice}
\end{figure}

\begin{figure}
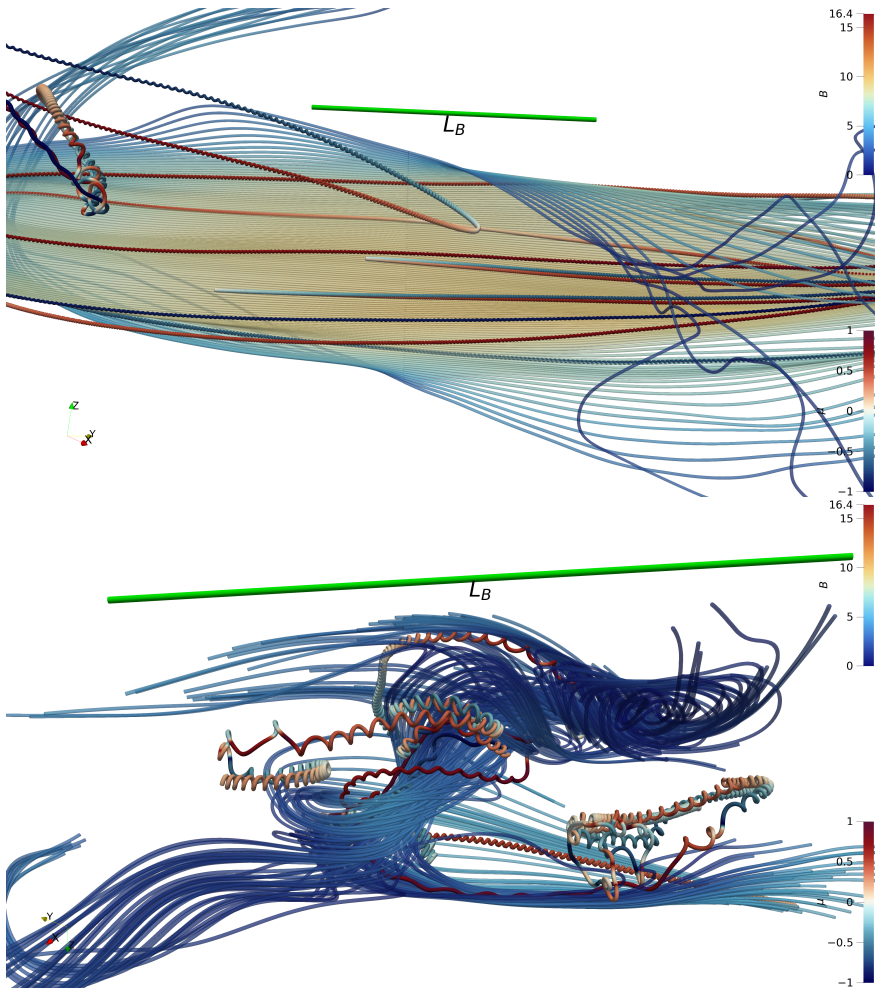

    \centering
    \includegraphics[width=.85\linewidth]{fig4a} \\
    \includegraphics[width=.85\linewidth]{fig4b}
    \caption{Magnetic field lines coloured by field strength~$B$ and test-particle trajectories coloured by pitch angle cosine~$\mu$ in the flux tube (\textit{top}) and one of the plasmoids (\textit{bottom}) from figure~\ref{fig:slice}.
    In the coherent flux tube, particles are closely bound to field lines with occasional large-scale mirroring.
    The plasmoid exhibits highly tangled field lines and effectively confines particles with a mixture of small-scale mirroring and unmagnetised scattering.
    The magnetic correlation scale is indicated for reference.}
    \label{fig:trajectories}
\end{figure}

At the outermost zoom level in figure~\ref{fig:iso}, we recognise that most of the magnetic energy is distributed intermittently throughout the domain, concentrated in a few intense flux tubes with almost circular cross-sections, which extend up to the flow correlation scale~$l\lesssim L_u$.
These structures are sometimes observed wrapped up by intense current sheets.
Further zooming into apparently quiet regions reveals a population of smaller, less intense, flattened flux tubes, organised in a tightly folded pattern~\citep{Schekochihin2004a} with embedded current sheets.
The networks of flux tubes and current sheets are dual to each other.
Outside of coherent flux tubes, e.g.~in the debris of disrupted structures, field lines tend to be chaotically tangled and highly disorganised.
Additionally, reconnection (either due to resistive diffusion or due to tearing) of folded flux tubes is observed to seed small-scale plasmoids~\citep{Galishnikova2022}.
This diverse picture is illustrated by a slice plot through an intense flux tube and its surroundings in figure~\ref{fig:slice}.

The conventionally computed correlation wavenumber~$k_{B,\mathrm{corr}}$ does not adequately reflect the geometric structure of the magnetic field, as indicated by its rather large value~$O(10\,k_\mathrm{forcing})$, because the intermittent and highly correlated flux tubes are averaged out.
It is thus instructive to also report the characteristic geometric scales~$k_\parallel$,~$k_{\vb{B}\times\vb{j}}$ and~$k_{\vb{B}\cdot\vb{j}}$, which are sensitive to the local direction of the magnetic field \citep[see table~\ref{tab:scales};][]{Schekochihin2004a,Galishnikova2022}.
For~$h=2$, these values confirm elongated tube-like structures with length~$l_\parallel\approx L_u$ and an approximate aspect ratio of~$1:6:6$.
However, the flow~$\vb{u}$ exhibits a lower degree of intermittency compared with~$\vb{B}$, so the correlation wavenumber is more suitable to characterise its correlation structure. The small discrepancy between~$k_{u,\mathrm{corr}}$ and~$k_\mathrm{forcing}$ results from the presence of a broadband turbulent energy spectrum \citep{Seta2020}.

\subsection{Test particle results}\label{sec:simuresults2}
Figure~\ref{fig:trajectories} shows examples of test-particle trajectories.
Particles are clearly magnetised with~$\bar{\kappa}\bar{r}_g<1$ inside the intense flux tube, where they closely follow magnetic field lines.
Due to large-scale variations of the illustrated flux tube, particles undergo occasional mirroring events, which appear as sudden reversals of direction.
These rare reversals imply a long parallel MFP.
Outside of the flux tube, particle motion is unmagnetised with~$\bar{\kappa}\bar{r}_g>1$ and appears much more erratic, as the particles bounce chaotically and with a short MFP through the tangled field.
An additional interesting case is illustrated by the small-scale plasmoid, which appears to be an efficient device for trapping particles.
However, since they appear to have a sub-dominant contribution in our statistics, and due to numerical concerns addressed in Section~\ref{sec:discuss-structures}, we focus here on the modelling of magnetised and unmagnetised motion, and
leave a detailed study on the effect of plasmoids on particle transport for later work.

\begin{figure}
    \centering
	\begin{subfigure}{0.49\linewidth}
        \includegraphics[width=\linewidth]{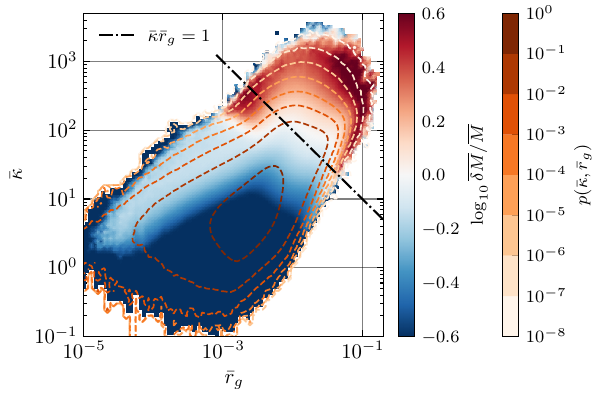}
		\caption{}
		\label{fig:cond-dMa}
	\end{subfigure}~%
	\begin{subfigure}{0.49\linewidth}
        \includegraphics[width=\linewidth]{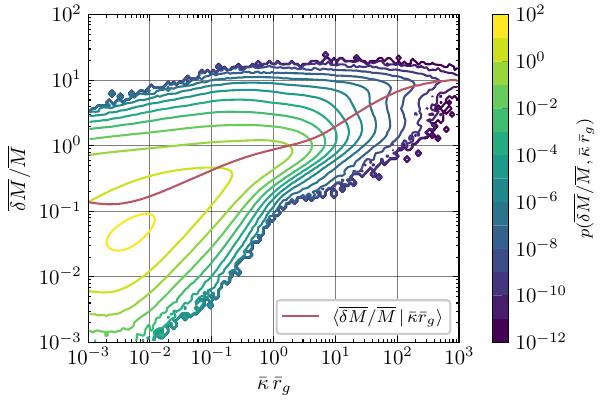}
		\caption{}
		\label{fig:cond-dMb}
	\end{subfigure}
    \caption{\textit{(a)}~Average of the relative magnetic moment variation~$\ol{\delta M}/\ol{M}$ conditional on particle gyro radius~$\bar{r}_g$ and field-line curvature~$\bar{\kappa}$.
    All recorded quantities are gyro-averaged.
    The colour scale is centred at~$\ol{\delta M}/\ol{M}=1$, where particles can be considered magnetised for smaller variations and unmagnetised for larger variations.
    Further, the colour scale is capped to~$\log_{10}\ol{\delta M}/\ol{M}\in(-0.6, 0.6)$ to highlight the transition region.
    This transition region is compared with the magnetisation criterion~$\bar{\kappa}\bar{r}_g\sim1$ expected from the field-line curvature picture.
    The joint density~$p(\bar{\kappa}, \bar{r}_g)$ is indicated for reference.
    \textit{(b)}~The conditional average~$\langle\ol{\delta M}/\ol{M}|\bar{\kappa}\bar{r}_g\rangle$ also shows the transition from predominantly magnetised and to predominantly unmagnetised motion as~$\bar{\kappa}\bar{r}_g$ increases, although the joint density~$p(\ol{\delta M}/\ol{M},\bar{\kappa}\bar{r}_g)$ reveals some uncertainty of this criterion.}
    \label{fig:cond-dM}
\end{figure}

\begin{figure}
    \centering
	\begin{subfigure}{0.49\linewidth}
        \includegraphics[width=\linewidth]{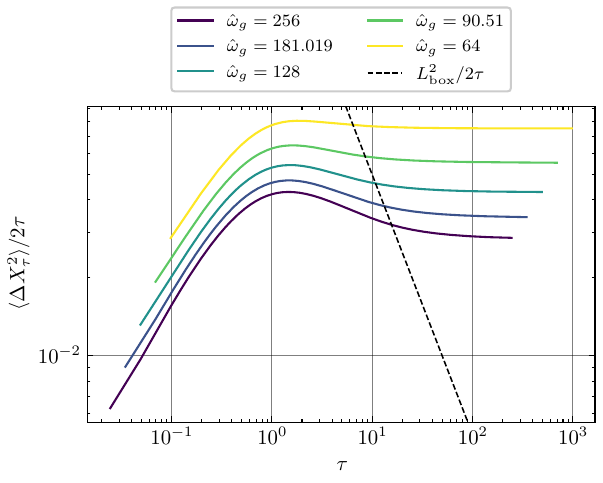}
		\caption{}
		\label{fig:rdca}
	\end{subfigure}~%
	\begin{subfigure}{0.49\linewidth}
        \includegraphics[width=\linewidth]{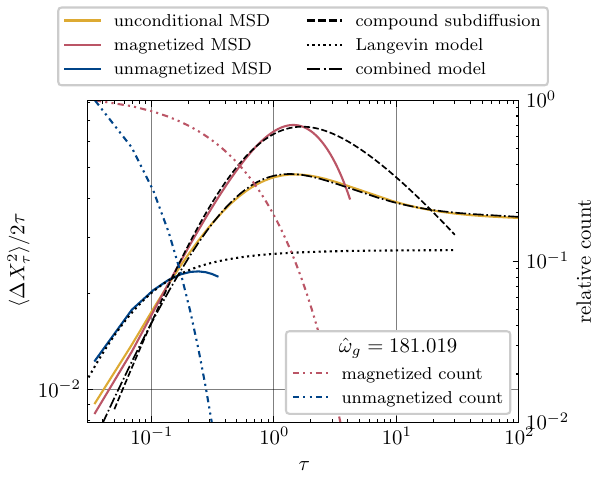}
		\caption{}
		\label{fig:rdcb}
	\end{subfigure}
    \caption{\textit{(a)}~MSD of the test particles, depending on the time lag~$\tau$, revealing initial ballistic propagation, transient subdiffusion and asymptotic diffusive behaviour. Notably, diffusion only occurs for mean square displacements beyond the simulation box size.
    \textit{(b)}~Conditional MSD for magnetised and unmagnetised transport for a selected test-particle energy, as well as the respective relative bin counts.
    Longer consecutively magnetised or unmagnetised segments have a lower probability than shorter ones, which leads to a systematically smaller sample size for the conditional averages at larger time scales~$\tau$.
    We describe magnetised transport by compound subdiffusion and unmagnetised transport by a Langevin equation.
    The discrepancy between magnetised data and model are likely due to this bias at large~$\tau$, which is addressed when tuning the combined model to the unconditional MSD, resulting in good agreement.}\label{fig:rdc}
\end{figure}

These observations are substantiated by the conditional statistics given by equations~(\ref{eq:dM}) and~(\ref{eq:dX}).
First, figure~\ref{fig:cond-dM} confirms that particles are magnetised in the sense that their magnetic moments are weakly conserved,~$\ol{\delta M}/\ol{M}<1$, when their gyro radii are smaller than the experienced field-line curvature radius, i.e.~$\bar{\kappa}\bar{r}_g<1$~\citep{Lemoine2023}.
If this condition is violated, i.e.~$\bar{\kappa}\bar{r}_g>1$, large variations of the magnetic moment are expected.
Although~$\bar{\kappa}\bar{r}_g=1$ does not constitute an exact decision boundary, we consider it as a useful heuristic, because our ensuing results remain qualitatively robust under minor refinements of this condition.
Appendix~\ref{app:perpcurv} shows results for an alternative magnetisation criterion based on the perpendicular reversal scale~$\kappa_\perp$ introduced by \cite{Kempski2023}, which can be approximated as~$\bar{\kappa}_\perp\bar{r}_g^{1/2}\sim30$.

Second, figure~\ref{fig:rdc} shows the unconditional MSD for our considered particles, as well as the conditional observations for~$\hat{\omega}_g=181.019$.
The unconditional data reveals three time scales of transport: particles move ballistically on short time scales, followed by a transient subdiffusive phase, before becoming asymptotically diffusive.
As noted in Section~\ref{sec:simusetup2}, convergence to normal diffusion occurs only after the MSD reaches the box size, which means that particles may experience the same structures repeatedly due to the periodic boundary conditions. 
We address this issue partly by modelling the transport behaviour on short and intermediate time scales, where the bias of the periodic domain is negligible.

The conditional data supports two distinct transport behaviours, where magnetised particles exhibit much longer MFPs compared to unmagnetised particles.
Also indicated are the relative bin counts contributing to the conditional averages.
This number strongly decreases for both cases for larger time scales, because the probability of finding an uninterrupted conditional trajectory segment decreases with its desired length.
The conditional averages at large time scales are thus likely dominated by a few very intense flux tubes, implying a systematic bias caused by our heuristic magnetisation criterion~$\bar{\kappa}\bar{r}_g\sim 1$.
Our methodology for modelling the conditional and unconditional data, as well as addressing this bias, is presented in Section~\ref{sec:model}.

\section{Stochastic transport model}\label{sec:model}
Motivated by our previous findings, we present a simplified stochastic model for the transport of charged test particles, resulting from a competition between magnetised and unmagnetised behaviour.
We assume that magnetised behaviour is dictated by the most intense flux tubes which extend up to the correlation scale of the turbulence~$\lambda_\mathrm{fl}\lesssim L_u$.
Particles are strictly bound to those field lines, but may undergo pitch-angle scattering with rate~$\theta_\mu$.
This scattering represents large-scale mirror interactions on the scale of the flux tubes, and thus, its MFP~$\lambda_\mu\sim c/\theta_\mu$ is expected to be long and comparable to the flux tube scale, i.e.~$\lambda_\mu\sim\lambda_\mathrm{fl}$.
Particles remain in the magnetised transport state for some mean duration~$t_\mu^\ast$, which represents the mean time until a scattering event due to sharp field-line curvature occurs.

However, unmagnetised behaviour results from an interplay between the intrinsic particle momentum and random scattering due to highly tangled field lines, where no well-defined mean field is seen by the particle, so we expect a relatively short MFP and high scattering rate~$\theta_\mathrm{scatter}\sim c/\lambda_\mathrm{scatter}$.
Particles wander around the domain for some mean duration~$t^\ast_\mathrm{scatter}$ until they encounter a strong flux tube and become magnetised again.

For the sake of comparison, we can estimate the MFP of the unconditional asymptotic diffusion process as~$\lambda_\mathrm{asymp}\sim D_\infty/c$ (this will be specified more precisely below), and we expect the ordering 
\begin{equation}
    \lambda_\mathrm{scatter}<\lambda_\mathrm{asymp}<\lambda_\mu\sim\lambda_\mathrm{fl}.
\end{equation}
The mean durations parametrise {escape-time probability distributions}~$p_\mu(t|t^\ast_\mu)$ and~$p_\mathrm{scatter}(t|t^\ast_\mathrm{scatter})$ for both transport modes.
They may be related to the distribution of field-line curvature, and from the dominance of low curvatures, we expect the ordering
\begin{equation}
    t^\ast_\mathrm{scatter}<t^\ast_\mu.
\end{equation}

We assume here that each kind of motion (specifically field-line wandering, magnetised mirroring and unmagnetised scattering) is sufficiently characterised by its respective MFP~$\lambda$ and thus adequately modelled by a Langevin equation, where both initial ballistic and asymptotic diffusive behaviour are resolved.
The emergent combined transport behaviour can then be studied as a competition between the involved MFPs and mean durations.
Further, the combination of two simple Langevin-like descriptions readily leads to compound subdiffusion and thus, a plausible explanation for the transient subdiffusion shown by the test particles.
We emphasise that qualitative descriptions are linked to the relative sizes of the various MFPs, for instance, ``coherent motion'' or ``large-scale mirroring'' correspond to~$\lambda_\mu\sim L_u$, while ``chaotic motion'' corresponds to~$\lambda_\mathrm{scatter}\ll L_u$.

\subsection{Stochastic differential equations}
We start with describing the motion of field lines and unmagnetised particles by a 3-D Langevin equation~\citep{Chandrasekhar1943,Bian2024}
\begin{equation}
    \dd{\vb{X}}_s=\vb{V}_s\dd{s},\quad
    \dd{\vb{V}}_s=-\theta\vb{V}_s\dd{s}+\sqrt{\frac{2}{3}\theta v^2_\mathrm{rms}}\dd{\vb{W}}_s,
    \label{eq:fieldline}
\end{equation}
where~$\vb{V}_s$ and~$\vb{X}_s$ denote velocity and position of the random walker at the generalised time or path parameter~$s$.
The walker is parametrised by the scattering rate~$\theta$ and thermal velocity~$v_\mathrm{rms}$, and driven by 3-D Gaussian white noise~$\dd{\vb{W}}_s$ with the correlation function
$\langle\dd{W}_{i,s}\dd{W}_{j,s'}\rangle=\delta_{ij}\delta(s-s')$.
The MSD is given by (as shown in Appendix~\ref{app:msd1})
\begin{equation}
    \langle X^2_s \rangle = \langle \|\vb{X}_s\|^2 \rangle = \frac{2v^2_\mathrm{rms}}{\theta^2}\left(e^{-\theta s}+\theta s-1\right),
    \label{eq:fieldline-dx2}
\end{equation}
which scales asymptotically as
$\langle X_s^2\rangle\underset{s\to0}{\sim}v_\mathrm{rms}^2s^2$ and
$\langle X_s^2\rangle\underset{s\to\infty}{\sim}{2v_\mathrm{rms}^2s}/{\theta}$.
The MFP is accordingly given by the scattering rate and thermal velocity as
\begin{equation}
    \lambda=\frac{v_\mathrm{rms}}{\theta}
    =\frac{D_{xx}}{v_\mathrm{rms}},
\end{equation}
with~$D_{xx}=\lim_{s\to\infty}{\langle X_s^2\rangle}/{2s}$.
In the following, we set~$v_\mathrm{rms}=1$ for field lines and unmagnetised transport in accordance with the normalisation choices~$\|\vb{V}\|=1$ and~$\langle\|\vb{B}\|^2\rangle=1$ according to Section~\ref{sec:simu}.

Next, we describe pitch-angle motion of magnetised particles with an Itô stochastic differential equation of the form~\citep{Strauss2017}
\begin{equation}
    \dd{s}_t=v_\mathrm{eff}\mu_t\dd{t},\quad
    \dd{\mu}_t=
    \frac{\dd{D_{\mu\mu}(\mu_t)}}{\dd{\mu}}\dd{t}
    +\sqrt{2D_{\mu\mu}(\mu_t)}\dd{W}_t,
    \label{eq:pitchanglewalker}
\end{equation}
where~$s_t$ denotes the displacement along a field line at time~$t$ and~$\mu_t\in(-1,1)$, with reflective boundary conditions, denotes the pitch angle cosine.
The effective velocity~$v_\mathrm{eff}$ reflects anisotropies of the pitch-angle-cosine distribution, where~$v_\mathrm{eff}=1$ indicates an isotropic uniform distribution.
The process is driven by uncorrelated Gaussian white noise with~$\langle\dd{W}_t\dd{W}_{t'}\rangle=\delta(t-t')$ and is characterised by the pitch angle diffusion coefficient
\begin{equation}
    D_{\mu\mu}(\mu)=D_0\left(1-\mu^2\right).
    \label{eq:Dmu}
\end{equation}
We choose this generic isotropic shape of~$D_{\mu\mu}$ in accordance with the agnostic stance of this work towards the detailed pitch-angle physics~\citep[see also][]{vandenBerg2024}. 
Analogously to equation~(\ref{eq:fieldline-dx2}), due to the Markovian nature of this process resulting in an exponential velocity correlation function, the MSD is given by
\begin{equation}
    \langle s_t^2\rangle=\frac{2\lambda_\mu^2}{3}\left(e^{-\theta_\mu t}+\theta_\mu t-1\right),
    \label{eq:pitchangle-dx2}
\end{equation}
where the additional factor~$1/3$ comes from the variance of the uniform distribution~$\mathcal{U}(-1, 1)$.
The MFP~$\lambda_\mu=v_\mathrm{eff}/\theta_\mu$ of this process is known from the literature~\citep{Schlickeiser2002} as
\begin{equation}
    \lambda_\mu
    =\frac{3D_{ss}}{v_\mathrm{eff}}
    =\frac{3v_\mathrm{eff}}{8}\int_{-1}^{+1}\frac{\left(1-\mu^2\right)^2}{D_{\mu\mu}(\mu)}\dd{\mu},
\end{equation}
with~$D_{ss}=\lim_{t\to\infty}\langle s_t^2\rangle/2t$, where the integral evaluates together with equation~(\ref{eq:Dmu}) to~$\lambda_\mu=v_\mathrm{eff}/2D_0$.

Further, a pitch-angle walker diffusing along a diffusing field line results in compound subdiffusion, with the MSD scaling as~$t^{1/2}$.
The spatial position of such a random walker is found by evaluating the field-line trajectory~$\vb{X}_{\mathrm{fl},s}$, given by equation~(\ref{eq:fieldline}), at the pitch-angle displacement coordinate~$s_t$, given by equation~(\ref{eq:pitchanglewalker}), as
\begin{equation}
    \vb{Y}_t=\vb{X}_{\mathrm{fl},s_t}.
    \label{eq:compound}
\end{equation}
This approach resembles Brownian yet non-Gaussian diffusion via subordination as presented by \cite{Chechkin2017}, and guided by their techniques, we can write the MSD as the integral
\begin{equation}
    \left\langle Y_{t}^2\right\rangle=\int_{-\infty}^{+\infty}\left\langle X_{\mathrm{fl},|s|}^2\right\rangle \frac{1}{\sqrt{2\upi\langle s_t^2\rangle}} \exp\left(-\frac{s^2}{2\langle s_t^2\rangle}\right)\dd{s},
    \label{eq:compound-dx2}
\end{equation}
where the MSD of the field line~$\langle X_{\mathrm{fl},s}^2\rangle$ is given by equation~(\ref{eq:fieldline-dx2}) and the MSD of the pitch-angle walker~$\langle s_t^2\rangle$ is given by equation~(\ref{eq:pitchangle-dx2}).
For the purpose of fitting equation~(\ref{eq:compound-dx2}) to the data, we evaluate the integral numerically with an adaptive Gaussian quadrature rule; however, the asymptotic behaviour can be written explicitly as~$\langle Y^2_{s_t}\rangle\underset{t\to0}{\sim}v_\mathrm{eff}^2t^2/3$ and~$\langle Y^2_{s_t}\rangle\underset{t\to\infty}{\sim}4\lambda_\mathrm{fl}\sqrt{v_\mathrm{eff}\lambda_\mu t}/3$,
as shown in Appendix~\ref{app:msd3}.

Based on these building blocks, we construct the combined stochastic model~$\vb{Z}_{\mathrm{model},t}$.
A stochastic walker alternates between magnetised and unmagnetised transport behaviour, where the duration for each segment is sampled from the respective escape-time probability distributions~$p(t|t^\ast_\mu)$ and~$p(t|t^\ast_\mathrm{scatter})$.
For magnetised transport, we simulate a field line~$\vb{X}_{\mathrm{fl},s}$ according to equation~(\ref{eq:fieldline}) and let the walker diffuse with~$s_t$ along this field line according to equation~(\ref{eq:pitchanglewalker}).
This behaviour is parametrised by the field-line and mirror MFPs~$\lambda_\mathrm{fl}$ and~$\lambda_\mu$, the effective velocity~$v_\mathrm{eff}$ and the magnetised mean duration~$t_\mu^\ast$.
Note that we simulate a new independent field line for every time the walker enters magnetised behaviour.
For unmagnetised transport, we simply simulate random scattering~$\vb{X}_{\mathrm{scatter},t}$ according to equation~(\ref{eq:fieldline}), which is parametrised by the scattering MFP~$\lambda_\mathrm{scatter}$ and the scattering mean duration~$t_\mathrm{scatter}^\ast$.
The procedure is summarised in Algorithm~\ref{alg:combined}.

\begin{figure}
\begin{algorithm}[H]
    \caption{Combined Stochastic Model}\label{alg:combined}
    \begin{algorithmic}[1]
        \Require number of steps $n$
        \State Choose initial magnetisation $m\in\{0,1\}$ \Comment{0: magnetised, 1: unmagnetised}
        \State Let $\vb{Z}_\mathrm{res}\gets[\vb{0}]$
        \While{$\Call{len}{\vb{Z}_\mathrm{res}}<n$}
        \If{$m=0$}
        \State Sample $t'\sim p(t|t^\ast_\mu)$
        \State Simulate $s_t\gets\mu\Call{-process}{0, t'}$ \Comment{Eq.~(\ref{eq:pitchanglewalker}) with $\lambda_\mu$}
        \State Simulate $\vb{X}_{\mathrm{fl},s}\gets\Call{fieldline-process}{\textsc{min}(s_t), \textsc{max}(s_t)}$ \Comment{Eq.~(\ref{eq:fieldline}) with $\lambda_\mathrm{fl}$}
        \State Interpolate $\vb{Y}_{t}\gets({\vb{X}_{\mathrm{fl},s},s_t})$ \Comment{Eq.~(\ref{eq:compound})}
        \State Append $\vb{Z}_\mathrm{res}\gets[\vb{Z}_\mathrm{res},\vb{Y}_{t}]$
        \State Let $m\gets1$
        \ElsIf{$m=1$}
        \State Sample $t'\sim p(t|t^\ast_\mathrm{scatter})$
        \State Simulate $\vb{X}_\mathrm{scatter,t}\gets\Call{scatter-process}{0,t'}$ \Comment{Eq.~(\ref{eq:fieldline}) with $\lambda_\mathrm{scatter}$}
        \State Append $\vb{Z}_\mathrm{res}\gets[\vb{Z}_\mathrm{res},\vb{X}_{\mathrm{scatter},t}]$
        \State Let $m\gets0$
        \EndIf
        \EndWhile
        \State \Return $\vb{Z}_\mathrm{res}$
    \end{algorithmic}
\end{algorithm}
\end{figure}

\subsection{Fitting the model}\label{sec:model-fit}
As noted in Section~\ref{sec:simuresults2}, finding connected segments of particle trajectories, which satisfy the desired conditions~$\bar{\kappa}\bar{r}_g\lessgtr1$ becomes more difficult with increasing length of these segments.
The reason for this is partly physical due to the distribution of field-line curvature and partly methodical due to our magnetisation condition~$\bar{\kappa}\bar{r}_g\sim1$ being merely a heuristic.
Thus the conditional data~$\langle\Delta X_\tau^2|\bar{\kappa}\bar{r}_g\lessgtr 1\rangle$ comes with the caveat, that data points at larger time scales~$\tau$ are backed by a much smaller sample size compared to data points at smaller~$\tau$.
This is illustrated by the relative bin counts in figure~\ref{fig:rdcb}.
When fitting our models, we address this issue by truncating the data at a cut-off time scale~$\tau_\mathrm{cutoff}$, where the choice of~$\tau_\mathrm{cutoff}$ should be small enough for sufficient statistical significance and large enough to provide a meaningful fit result.
For the magnetised case, we choose~$\tau_\mathrm{cutoff}=1.5\tau_\mathrm{peak}$ with~$\tau_\mathrm{peak}=\mathop{\mathrm{argmax}}_\tau\langle\Delta X^2_\tau|\bar{\kappa}\bar{r}_g<1\rangle/2t$ and ensure a minimum relative bin count of~$1\%$.
Analogously for the unmagnetised case, we choose~$\tau_\mathrm{cutoff}=1\tau_\mathrm{peak}$ with~$\tau_\mathrm{peak}=\mathop{\mathrm{argmax}}_\tau\langle\Delta X^2_\tau|\bar{\kappa}\bar{r}_g>1\rangle/2t$ and ensure a minimum relative bin count of~$1\%$ as well.
Despite these precautions, a systematic bias likely remains in the data, most notably in the conditional magnetised data, which is dominated at long time scales by a small number of very intense flux tubes.
This remaining bias is addressed by the Bayesian optimisation of the magnetised mean duration~$t^\ast_\mu$ and MFP~$\lambda_\mu$ discussed below.

\begin{figure}
    \centering
	\begin{subfigure}{0.49\linewidth}
        \includegraphics[width=\linewidth]{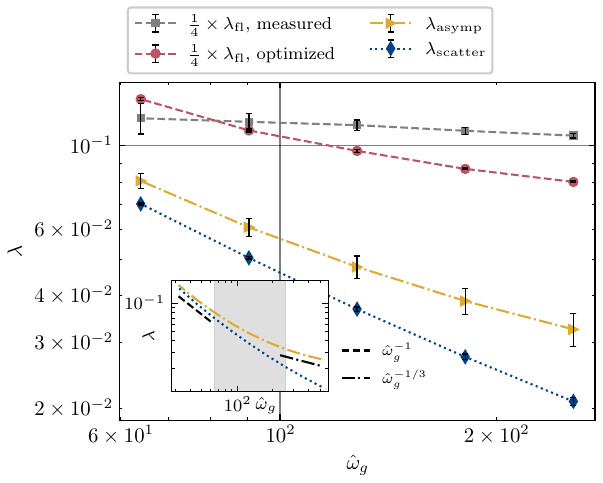}
		\caption{}
		\label{fig:mfpa}
	\end{subfigure}~%
	\begin{subfigure}{0.49\linewidth}
        \includegraphics[width=\linewidth]{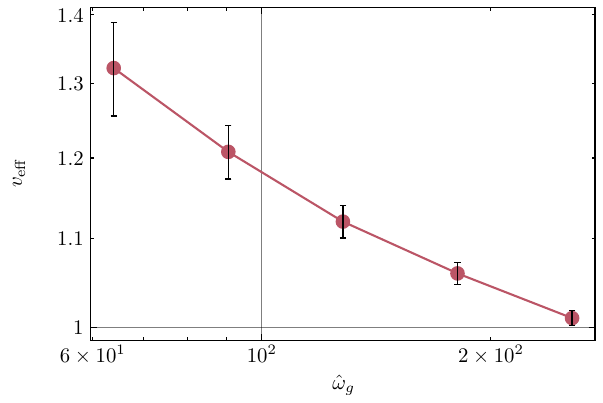}
		\caption{}
		\label{fig:mfpb}
	\end{subfigure}
    \caption{\textit{(a)}~Fitted MFPs for conditional magnetised~$\lambda_\mathrm{fl}$ and unmagnetised transport~$\lambda_\mathrm{scatter}$, as well as for the unconditional asymptotic case~$\lambda_\mathrm{asymp}$.
    As shown in the inset,~$\lambda_\mathrm{asymp}$ converges to~$\lambda_\mathrm{scatter}$ for high energies, where our magnetised model is no longer valid.
    Scalings~$\hat{\omega}_g^{-1}$ and~$\hat{\omega}_g^{-1/3}$ are indicated for reference.
    The field line MFP~$\lambda_\mathrm{fl}$ is obtained twice: once by naively fitting equation~(\ref{eq:compound-dx2}) to the biased magnetised MSD \textit{(measured)} and once by optimising the unbiased loss function given by equation~(\ref{eq:loss}) \textit{(optimised)}.
    Both values are scaled by a factor~$\frac{1}{4}$ to simplify comparison with~$\lambda_\mathrm{scatter}$ and~$\lambda_\mathrm{asymp}$.
    \textit{(b)}~Fitted effective velocity of magnetised pitch-angle diffusion.
    The error bars in both plots for the fitted models are given by $1.96\ \times$~the standard error produced by the respective fit routines.
    The error bars for~$\lambda_\mathrm{asymp}$ are obtained by taking the mean and $1.96\ \times$~the standard deviation over the eight independent MHD snapshots.}
    \label{fig:mfp}
\end{figure}

We proceed by fitting the compound subdiffusion model given by equation~(\ref{eq:compound-dx2}), which is parametrised by~$\lambda_\mathrm{fl}$,~$\lambda_\mu$ and~$v_\mathrm{eff}$, to the magnetised data~$\langle\Delta X^2_\tau|\bar{\kappa}\bar{r}_g<1\rangle$.
Since a clear distinction of the two MFPs requires reliable data at large~$\tau$, we argue that large-scale magnetic mirrors mostly operate on the correlation scale of coherent flux tubes and fix~$\lambda_\mathrm{fl}=\lambda_\mu$, thereby reducing the number of free parameters by one.
Further, we fit the Langevin model given by equation~(\ref{eq:fieldline-dx2}), which is parametrised by~$\lambda_\mathrm{scatter}$, to the unmagnetised data~$\langle\Delta X_\tau^2|\bar{\kappa}\bar{r}_g>1\rangle$.
We employ a nonlinear least squares method for both cases (\citealt{Branch1999}).
For comparison, the unconditional asymptotic MFP of the unconditional data is estimated as
\begin{equation}
    \lambda_\mathrm{asymp}=\frac{D_\infty}{v_\mathrm{rms}}
\end{equation}
with the asymptotic diffusion coefficient~$D_\infty=\lim_{\tau\to\infty}\langle\Delta X_\tau^2\rangle/2\tau$ and the particle r.m.s.~velocity~$v_\mathrm{rms}^2=\frac{3}{2}\lim_{\tau\to0}\langle\Delta X_\tau^2\rangle/\tau^2$.
This choice of~$v_\mathrm{rms}$ ensures consistency with the Langevin model for the unmagnetised motion.
The resulting MFPs are shown in figure~\ref{fig:mfpa} and discussed in Section~\ref{sec:model-results}.
The effective pitch angle velocities~$v_\mathrm{eff}$ are shown in figure~\ref{fig:mfpb}, where~$v_\mathrm{eff}=1$ corresponds to an isotropic uniform pitch angle cosine distribution.
We observe~$v_\mathrm{eff}>1$ for higher particle energies, which is likely due to shorter durations of magnetised segments, thus those particles experience less mirroring, which would isotropise the pitch angles.

\begin{figure}
    \centering
	\begin{subfigure}{0.49\linewidth}
        \includegraphics[width=\linewidth]{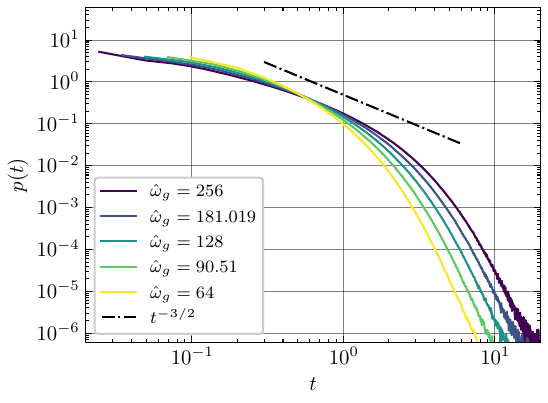}
        \caption{}
		\label{fig:etda}
	\end{subfigure}~%
	\begin{subfigure}{0.49\linewidth}
        \includegraphics[width=\linewidth]{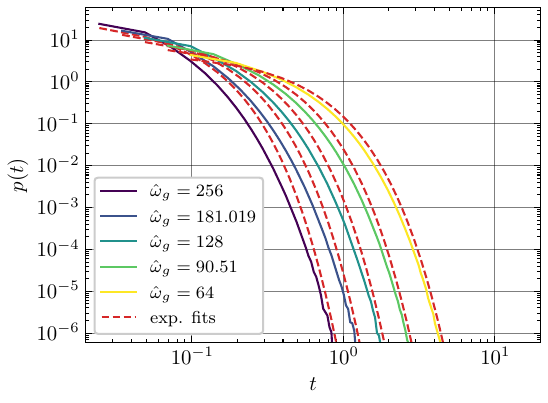}
        \caption{}
		\label{fig:etdb}
	\end{subfigure}
    \caption{Escape-time probability distributions estimated from the conditional test-particle averages.
    \textit{(a)}~Magnetised cases exhibit heavier tails than an exponential distribution, indicating that the magnetised motion has some memory.
    The power-law scaling~$t^{-3/2}$ expected for the classical first-passage time distribution of a random walker on a finite line is indicated for reference.
    \textit{(b)}~Unmagnetised cases closely resemble exponential distributions, indicating a memory-less Markov nature of the unmagnetised motion.}
    \label{fig:etd}
\end{figure}

\begin{figure}
    \centering
	\begin{subfigure}{0.49\linewidth}
        \includegraphics[width=\linewidth]{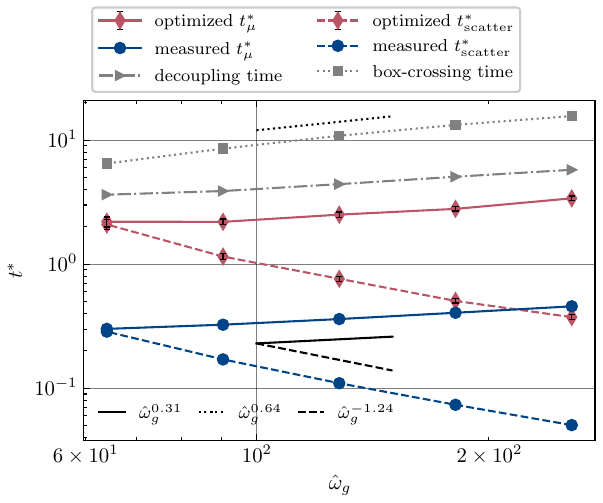}
        \caption{}
		\label{fig:timesa}
	\end{subfigure}~%
	\begin{subfigure}{0.49\linewidth}
        \includegraphics[width=\linewidth]{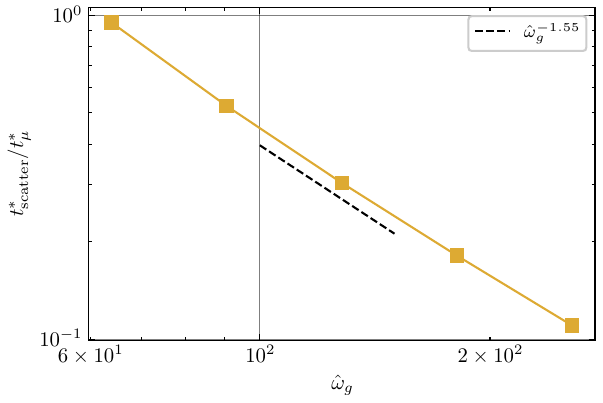}
        \caption{}
		\label{fig:timesb}
	\end{subfigure}
    \caption{\textit{(a)}~Measured and fitted mean durations~$t^\ast_\mu$ and~$t^\ast_\mathrm{scatter}$. Box-crossing and decoupling time scales are given for reference.
    The measured magnetised mean duration is much smaller than the expected decoupling time scale due to neglecting the correlation of large-scale flux tubes.
    However, the optimised mean magnetised duration is comparable to the decoupling time scale. 
    The error bars of the optimised results are given by $1.96\ \times$~the estimated confidence interval of the Bayesian optimisation procedure (see Appendix~\ref{app:bayes}).
    \textit{(b)}~Ratio of measured unmagnetised and magnetised mean durations, as a proxy for the volume filling fraction of scattering sites experienced by the test particles. The optimised mean scattering duration is determined by multiplying this measured ratio with the optimised mean magnetised duration.}
    \label{fig:times}
\end{figure}

Next, we estimate the escape-time probability distributions by recording the lengths of observed conditional magnetised and unmagnetised segments into histograms.
The densities of the histogram are plotted in figure~\ref{fig:etd}, revealing an exponential shape for the unmagnetised case, which corroborates the view of the unmagnetised motion as a memory-less random walk.
On the other hand, the magnetised case roughly resembles a tempered power-law shape, indicating a process with long memory governed by extended flux tubes.
When picturing a 1D pitch-angle walker on a field line with finite length, one would expect the classical first-passage time distribution~\citep{Krapivsky2010}, which scales as~$t^{-3/2}$. This scaling is indicated for reference.

To judge whether the results are reasonable, we compare the magnetised mean duration~$t_\mu^\ast$ with the typical box-crossing time scale~$\tau_\mathrm{box}$ (see figure~\ref{fig:rdca}) and the decoupling time scale~$\tau_\mathrm{decouple}$, which is estimated as the inflection point of the unconditional running diffusion coefficient~$\langle\Delta X_\tau^2\rangle/2\tau$ as
\begin{equation}
    \tau_\mathrm{decouple}=\mathop{\mathrm{argmin}}_{\tau\in(0,\infty)}\frac{\dd{}\log{\langle\Delta X_\tau^2\rangle}/{2\tau}}{\dd{}\log \tau}.
\end{equation}
First, since the dominant coherent structures which govern the particle transport are comparable to the fluid correlation scale~$l\lesssim L_u<L_\mathrm{box}$, we expect~$t_\mu^\ast<\tau_\mathrm{box}$, i.e. that magnetised motion is clearly separated from the box-crossing time scale.
Second, we assume that the initially ballistic and transiently subdiffusive behaviour of the unconditional MSD is given by magnetised transport, and asymptotic diffusion is linked to decoupling of particles from coherent field lines, so we expect~$t^\ast_\mu\sim\tau_\mathrm{decouple}$.
The various time scales are shown in figure~\ref{fig:timesa}, which confirms the first expectations, but reveals that the measured magnetised mean durations are shorter than expected by an order of magnitude~$\tau_\mathrm{decouple}/t_\mu^\ast\sim O(10)$.
Additionally, the combined model given by Algorithm~\ref{alg:combined} does not reproduce the unconditional MSD with these naively measured parameters (not shown).

This severe underestimation is likely caused by the discrepancy between our combined stochastic model, which simulates a new independent stochastic field line for each magnetised segment, and test particles in MHD snapshots, where magnetised transport is governed by a finite number of spatially correlated intense flux tubes.
To explore the capabilities of our combined model, we search for an optimal effective mean magnetised duration~$t^\ast_\mu$ by means of Bayesian optimisation (see Appendix~\ref{app:bayes} for details).
Specifically, we minimise the loss function
\begin{equation}
    \mathcal{L}(t^\ast_\mu,\lambda_\mathrm{fl})=
    \max_{\tau\in(0,\tau_\mathrm{max})}\left|\log\frac{\langle\Delta X^2_\tau\rangle}{\langle Z_{\mathrm{model},\tau}^2|t^\ast_\mu,\lambda_\mathrm{fl}\rangle}\right|,
    \label{eq:loss}
\end{equation}
which compares the unconditional MSD of test particles~$\langle\Delta X_\tau^2\rangle$ with the MSD of our combined model~$\langle Z^2_{\mathrm{model},\tau}|t^\ast_\mu,\lambda_\mathrm{fl}\rangle$.
We also take the field line MFP~$\lambda_\mathrm{fl}$ as a tuneable parameter to account for the previously discussed bias of the magnetised data.
Further, the effective pitch-angle walker velocity~$v_\mathrm{eff}$ and scattering MFP~$\lambda_\mathrm{scatter}$ are taken from the conditional fit results, the pitch angle MFP is fixed as~$\lambda_\mu=\lambda_\mathrm{fl}$ and the unmagnetised mean duration is determined from the fixed measured ratio~$t^\ast_\mathrm{scatter}/t^\ast_\mu$ (see figure~\ref{fig:timesb}), which serves as a proxy for the volume filling fraction of scattering sites.

\subsection{Model results}\label{sec:model-results}

\begin{figure}
    \centering
    \includegraphics[width=.9\linewidth]{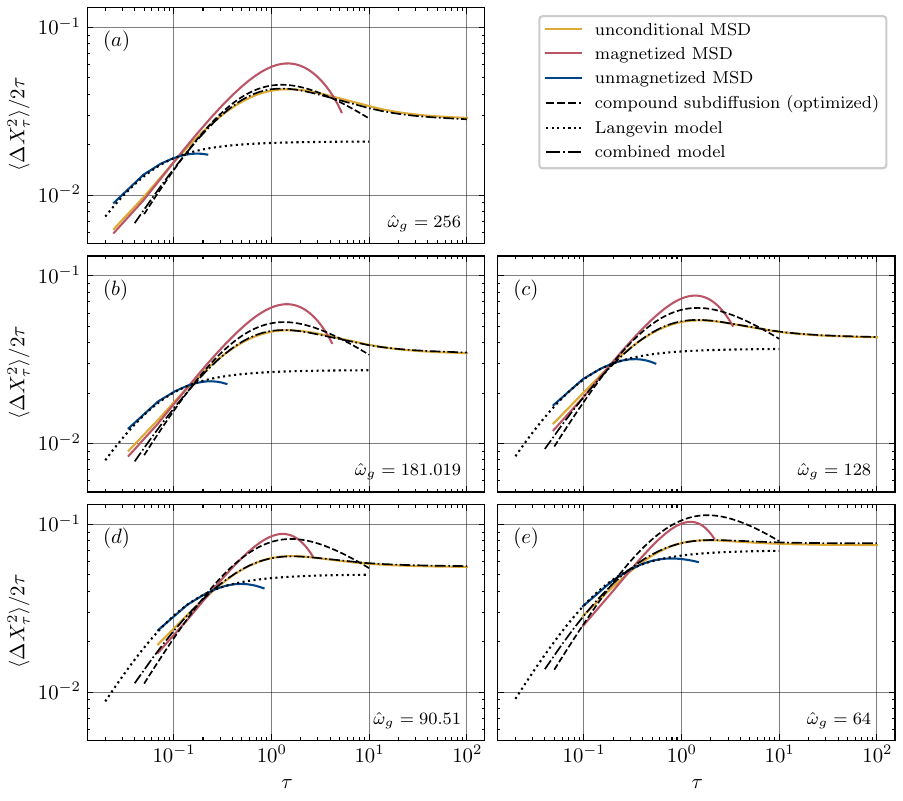}
    \caption{Time-dependent MSDs for conditional and unconditional test-particle measurements, as well as conditional and combined model results.
    The combined model with optimised parameters shows good agreement with the unconditional test-particle MSD.
    The compound subdiffusion model is also shown for the unbiased optimised parameters, thus some deviation from the biased conditional data is present.}
    \label{fig:results}
\end{figure}

The MSDs of the combined model for the optimised values are plotted in figure~\ref{fig:results} and agree well with the unconditional test-particle MSD for all considered particle energies.
Note that the optimised magnetised mean durations are now comparable to the decoupling time scale, as shown in figure~\ref{fig:timesa}.
Also plotted in figure~\ref{fig:results} are the conditional MSDs and models, showing good agreement of the simple Langevin diffusion with the unmagnetised data, as well as some disagreement for the compound subdiffusion and the biased magnetised data.
Specifically, as shown in figure~\ref{fig:mfpa}, the optimisation procedure predicts for most particle energies smaller MFPs compared to the conditional data.
These smaller values can account for transport by less intense flux tubes or constellations of strongly correlated flux tubes and, thus, correct the bias from the conditional magnetised data, dominated by isolated (intrinsic to the measurement methodology) and very intense flux tubes (fulfilling the heuristic magnetisation criterion for long time scales).

\section{Discussion}\label{sec:discussion}
\subsection{Implications on cosmic-ray transport}\label{sec:discussion-asymptotic}

We emphasise that the considered regime of particle energies is non-asymptotic and limited by the available numerical resolution of the MHD simulation.
However, in our results, consisting of MFPs in figure~\ref{fig:mfpa} and mean durations in figure~\ref{fig:timesa}, tendencies towards the asymptotic high- and low-energy regimes can be recognised.
On one hand, the high-energy regime with~$\hat{\omega}_g\lesssim1$ will be entirely governed by unmagnetised motion, because particles with gyro radii~$r_g\gtrsim L_u$ average out any non-trivial structures, as shown by increasing~$t^\ast_\mathrm{scatter}/t^\ast_\mu$ and~$\lambda_\mathrm{asymp}/\lambda_\mathrm{scatter}\to1$ as~$\hat{\omega}_g$ decreases.
The expected random-walk scaling is~$\lambda\sim\hat{\omega}_g^{-2}$ (\citealt{Reichherzer2020}; the indicated scaling~$\lambda\sim\hat{\omega}_g^{-1}$ in figure~\ref{fig:mfpa} is only transitional, compare also \citealt{Lubke2024}).
The conditional magnetised model is valid to roughly~$\hat{\omega}_g\sim64$.

On the other hand, the low-energy regime~$\hat{\omega}_g\gg 1$ becomes increasingly dominated by magnetised motion as indicated by decreasing~$t^\ast_\mathrm{scatter}/t^\ast_\mu$. The asymptotic MFP~$\lambda_\mathrm{asymp}$ likely becomes a function of the coherent field line MFP~$\lambda_\mathrm{fl}$ and the large-scale mirror MFP~$\lambda_\mu$ (our previous assumption~$\lambda_\mathrm{fl}=\lambda_\mu$ is not expected to hold in general), as well as the mean magnetised duration~$t^\ast_\mu$, which indicates the typical particle time scale of decoupling from field lines.
The scaling~$\lambda\sim\hat{\omega}_g^{-1/3}$ indicated in figure~\ref{fig:mfpa} serves purely as reference and should be taken with care, as small gyro radii with~$\hat{\omega}_g\gtrsim256$ are significantly affected by the MHD dissipation scales (compare also figure~\ref{fig:mhdspectrum}).

Our results clearly highlight the crucial role of coherent magnetic structures, especially for the low-energy regime, and support the role of field-line wandering in cosmic ray transport, as discussed by \citet{Pezzi2024}.
Further, highly magnetised low-energy particles such as GeV galactic cosmic rays are expected to be dynamically relevant on the plasma dynamics, requiring the consideration of streaming \citep{Sampson2023}.

To speculate about the asymptotic behaviour of the parameters of magnetised motion, we presume that the field line MFP~$\lambda_\mathrm{fl}$ for a particle with~$\hat{\omega}_g$ is given by the average coherent curvature radius, i.e. the average of curvature radii larger than the typical gyro radius~$\frac{\pi}{2}\hat{\omega}_g^{-1}$,
\begin{equation}
    \lambda_\mathrm{fl}(\hat{\omega}_g)=\left[\int_0^{\kappa_c}p(\hat{\kappa})\dd{\hat{\kappa}}\right]^{-1}\int_0^{\kappa_c}\frac{p(\hat{\kappa})\dd{\hat{\kappa}}}{\hat{\kappa}}
    \label{eq:curv-mfp}
\end{equation}
with $\kappa_c=\frac{2}{\upi}\hat{\omega}_g$.
Following \citet{Lemoine2023}, we account for the influence of the magnetic field strength on the particle gyro radius by considering the compensated field-line curvature~$\hat{\kappa}=\kappa/B$, whose distribution scales as~$p(\hat{\kappa})\sim\hat{\kappa}^{-2}$ \citep[see also][]{Baggaley2010,Yang2019,Bandyopadhyay2020}.
To compute equation~(\ref{eq:curv-mfp}), we assume the model distribution \citep[motivated by][]{Schekochihin2002b}
\begin{equation}
    p(\hat{\kappa})=s(s-1)\kappa_0^{s-1}\frac{\hat{\kappa}}{(\kappa_0+\hat{\kappa})^{s+1}},
\end{equation}
which agrees well with our MHD simulation for~$s=2$ and~$\kappa_0=1.5$ and yields $\lim_{\hat{\omega}_g\to\infty}\lambda_\mathrm{fl}(\hat{\omega}_g)=\kappa_0^{-1}$.

Further, the path lengths between large-scale mirror events likely follow a broad distribution, which is truncated by the typical length of coherent field lines, so we remain with our assumption~$\lambda_\mu\approx\lambda_\mathrm{fl}$ for now.
Given constant field-line and mirroring MFPs~$\lambda_\mathrm{fl}\approx\lambda_\mu\approx\kappa_0^{-1}$ in the low-energy limit, the asymptotic MFP is then solely determined by the decoupling time scale~$t^\ast_\mu$, which, in combination with compound subdiffusion of particles along coherent field lines, predicts~$\lambda_\mathrm{asymp}\sim {t^\ast_\mu}^{-1/2}$.
As shown in figure~\ref{fig:etda}, the durations of magnetised motion follow a broad power-law distribution with a long-time cut-off, which increases with decreasing particle energy.
The asymptotic scaling of~$t^\ast_\mu$ is not resolved in our data and can only be speculated about at the moment.
For instance,~$\lambda_\mathrm{asymp}\sim\hat{\omega}_g^{-1/3}$ would suggest~$t^\ast_\mu\sim\hat{\omega}_g^{2/3}$.
We expect it to arise from a complex interplay of lengths of coherent structures~$\lambda_\mathrm{fl}$, mirroring rates~$c/\lambda_\mu$, drift motions and possibly gyro-resonance due to Alfvén waves travelling along coherent flux tubes.

Further, as noted in Section~\ref{sec:simusetup2}, the reported values for~$\lambda_\mathrm{asymp}=D_\infty/v_\mathrm{rms}$ in figure~\ref{fig:mfpa} are biased by the periodic boundary conditions of the simulation domain (compare figure~\ref{fig:rdca}).
Despite this bias our considerations clearly show that the value of~$D_\infty$ emerges from the short-time evolution of the particle distribution which is influenced by coherent structures on scales below the box size.
Notably, these ``structure-mediated'' values are larger than those obtained from random-phase turbulence~\citep[see also][]{Shukurov2017,Lubke2024}.
However, on large scales~$\gg L_\mathrm{box}$, such as the ISM or ICM outer scales, this structure-mediated enhancement has to be weighted against confinement in structures if scattering is rare and~$t^\ast_\mu$ is large.

\subsection{Discussion of coherent structures}\label{sec:discuss-structures}
Despite the idealised nature of our turbulence set-up, consisting of a visco-resistive incompressible MHD simulation with~$Re_T\sim O(100)$ and~$Pr={\nu}/{\eta}=1$, our results can be contextualised with coherent structures ubiquitously found in simulations and observations~\citep[see, e.g.][]{Robitaille2020,Ntormousi2024}.
The fluctuating dynamo with compressible~\citep{Beattie2024} or kinetic~\citep{St-Onge2018} physics also exhibits pronounced coherent structures, although their statistics, such as~$p(B,\kappa)$, likely differ from our case and a dedicated study of cosmic-ray transport in these cases would be useful.
While our focus has been on intense flux tubes generated by the fluctuating dynamo,
coherent structures also emerge as patch-wise aligned states due to selective decay~(\citealt{Matthaeus2015,Hosking2021}; see also figure~\ref{fig:slice}) or as intense aligned wave packets in critically balanced turbulence~\citep{Perez2009,Chandran2015}.
Generally, different turbulence set-ups (e.g. forced or freely decaying, with or without an imposed background magnetic field, different degrees of compressibility, cross- and magnetic helicity) lead to different predominant kinds of structures, which may impact cosmic-ray transport in different ways.
Here, we use the term ``structure'' to loosely summarise relatively coherent phenomena with nonlinear qualities~(see also \citealt{Groselj2019} for a different, but not contradictory, point of view).

Current sheets emerge as the unifying feature of these various kinds of turbulence~\citep[see, e.g.][]{Servidio2011,Zhdankin2016},
which, given sufficient dynamical range, become tearing-unstable, leading to fast reconnection and the seeding of small-scale plasmoid chains, i.e. complexes of highly-tangled small-scale flux tubes.
This process is associated with a reconnection-mediated turbulence regime at the transition to kinetic scales~\citep{Loureiro2017,Mallet2017,Chernoglazov2021,Dong2022},
which has important implications on cosmic-ray acceleration~\citep{Comisso2019} and may be relevant for a transport theory of the lowest-energy cosmic rays or suprathermal particle populations.
However, particular care must be taken when analysing simulations featuring plasmoids, such as ours, because they may also appear as numerical artefacts for insufficient numerical resolution~\citep{Morillo2025}.

Understanding these various structures and their interaction with charged particles is crucial for a comprehensive picture of cosmic-ray transport in realistic multiphase media~\citep{Armillotta2022,Beattie2025b}.
Also, given sufficiently high numerical resolution, gyro-resonance can be expected to re-appear, for instance, mediated by small-scale Alfvén wave turbulence emerging along coherent flux tubes, either as part of the turbulent cascade or self-generated by cosmic rays.
A new cosmic-ray transport theory should carefully weigh these different processes and their varying predominance against each other~\citep{Kempski2022,Hopkins2022a}.

\subsection{Synthetic turbulence}
Synthetic turbulence refers to a class of algorithms designed for overcoming the limited resolvable range of scales of first-principles simulations.
They provide fast generation of random fields resembling certain statistics expected for realistic turbulence.
Most commonly, guided by the paradigm of gyro-resonance, these fields are synthesised from a prescribed energy spectrum and random Fourier phases~\citep[see, e.g.][]{Mertsch2020}.
These random-phase models have well-known shortcomings, such as the inability to reproduce intermittency and coherent structures, as illustrated by, e.g.~\cite{Shukurov2017}, which has led to the exploration of intermittent and structured models~\citep[][]{Subedi2014,Pucci2016,Durrive2022,Lubke2024,Maci2024,Lesaffre2025}.
To address the crucial geometric structure, future synthetic models for strong isotropic turbulence should include the field-line curvature distribution~$p(\kappa)\sim\kappa^{-5/2}$ as well as the anti-correlation between magnetic field strength and field-line curvature~$B\sim\kappa^{-1/2}$.
Instead of generating a synthetic field directly, Generative Diffusion Models, a recently developed machine learning technique, are able to learn and synthesise particle trajectories directly~\citep{Li2023,Martin2025}.
Such a technique could also be useful for classifying the various transport behaviours and refining the heuristic magnetisation criterion~$\bar{\kappa}\bar{r}_g\sim1$.

\section{Outlook}\label{sec:outlook}
\subsection{Refinement of the combined stochastic model}
We emphasise that our model is semi-empirical, where parameters are inferred from the test-particle data.
It is essential for a proper transport theory to predict the involved MFPs and mean durations from the underlying turbulence.
In particular, such a theory should account for the hierarchy of scales present in the problem, from scattering on gyro scales, over motion within individual coherent structures and up to large-scale network-like constellations of the population of coherent structures~\citep{Ntormousi2024}.
This task could be approached by projector-based coarse-graining of the dynamics, following the Mori-Zwanzig formalism~\citep{hudson-li:2020,lin-tian-etal:2021}.

Another possible approach to accurately derive the required MFPs and mean durations would be to follow \cite{drummond:1982} and start with a path integral formulation of turbulent diffusion. 
However, in Drummond's approach, the magnetic potential is assumed to be Gaussian.
A possible, far from trivial approach would be a path integral formulation including the action~$S_\mathrm{MHD}$ corresponding to the MHD dynamics plus the action~$S_\mathrm{particle}$ of the particle dynamics and applying non-perturbative methods to approximate this complex path integral~\citep[see, e.g.][]{grafke-grauer-schaefer:2015c,burekovic-schaefer-grauer:2024,schorlepp-grafke:2025}.

\subsection{Alternative transport descriptions}\label{sec:outlook-alternative}
A central assumption of our model is that all involved kinds of motion are Gaussian and sufficiently characterised by their MFPs or MSDs.
While our combined model of Langevin and compound subdiffusion successfully reproduces the test-particle MSD, possibly relevant higher-order statistics are neglected.
For instance, intermittent pitch-angle scattering~\citep{Zimbardo2020} or streaming~\citep{Sampson2023} can lead to superdiffusion of cosmic rays.
In the context of our test-particle data, magnetised motion may be described by spatial superdiffusion (averaged over the pitch angle), and mirror-confinement in coherent structures may correspond to subdiffusion.
A clever combination of Lévy walks and extended waiting times~\citep{Zaburdaev2015} with truncated distributions~\citep{Liang2025} may reproduce the intricate time evolution of the test-particle MSDs, while also accounting for possibly relevant higher-order statistics.
In this case of competition between sub- and superdiffusion~\citep{Magdziarz2007} of cosmic rays, a single (fractional) transport equation might be formulated that is equivalent to the stochastic model as presented for the superdiffusive case in \cite{Effenberger2024} and \cite{Aerdker2025}.

Alternatively, especially when modelling large-scale ($\gg L_u$) multiphase media, the structured nature of turbulence may be reflected by distinct diffusion coefficients.
For the example of two spatially distinguishable phases, one can employ a regular Fokker-Planck equation~$\partial_t f(\vb{x},t)=\Delta D(\vb{x})f(\vb{x},t)$ with a mixed diffusion coefficient~$D(\vb{x})=D_1$ for~$\vb{x}\in\text{phase 1}$ and~$D(\vb{x})=D_2$ for~$\vb{x}\in\text{phase 2}$, which is readily applicable to astrophysical problems \citep[see, e.g.][]{Reichherzer2025}.
Otherwise, a temporal switching Fokker-Planck equation~$\partial_tf_i(\vb{x},t)=D_i\Delta f_i(\vb{x},t)+A_{ij}f_j(\vb{x},t)$ with Markovian switching rates~$A_{ij}$ may be useful~\citep[see, e.g.][]{Bressloff2017,Balcerek2023}, which could be applied as a sub-grid model for cosmic-ray transport.

All presented models involve a competition between two transport regimes to capture the transient phases of the observed MSD. Similar switching or competing models are also known in other complex systems~\citep[see, e.g.][]{Doerries2022,Datta2024}.

\subsection{Generalised transport equation}
To study structure-mediated transport of cosmic rays in astrophysical applications, the physics described by our combined stochastic model needs to be encapsulated in a generalised transport equation for the $5+1$-dimensional particle distribution function~$f(\vb{x},\mu,\hat{\omega}_g,t)$.
This transport equation takes the form~$\partial_tf=\mathcal{D}[f]$, where the generalised transport operator~$\mathcal{D}[\cdot]$ can describe a wide range and combination of processes, such as (anomalous) spatial diffusion~$\mathcal{D}=D_{xx}\Delta^{(\alpha)}$,
pitch-angle diffusion~$\mathcal{D}=\partial_\mu D_{\mu\mu}\partial_\mu$,
or momentum diffusion~$\mathcal{D}=p^{-2}\partial_pp^2D_{pp}\partial_p$ (with momentum~$p\propto\hat{\omega}_g$ in our notation)~\citep{Metzler2000,Schlickeiser2002}.
Since transport properties are strong functions of the local plasma conditions and cosmic rays can exert a dynamically relevant feedback on the plasma, the transport equation needs to be coupled with the astrophysical simulation for a self-consistent treatment~\citep{Bai2015,Pfrommer2017,Boss2023}.
With such a set-up, one could, for instance, study the effect of streaming on coherent structures~\citep[compare, e.g.][]{Rieder2017,Sampson2023}.

Possible approaches to construct an operator~$\mathcal{D}$, which describes the competition between two modes of transport, are contemplated in Section~\ref{sec:outlook-alternative}.
We also note that~$\mathcal{D}$ must not necessarily be known explicitly, as it can be represented and solved by the corresponding stochastic differential equations~\citep[][]{Effenberger2024,Aerdker2025}.
The remaining challenge is then to parametrise our transport model in terms of characteristic turbulence quantities, which may vary across the large-scale simulation domain, such as the turbulence correlation scale, turbulence strength or sonic Mach number.

Such an integrated study can provide data for validating this or other non-standard transport models against observational data.
Ultimately, this study should not only be consistent with the data, but provide falsifiable predictions.
Observational constraints for intermittent cosmic-ray scattering theories are, for instance, considered by \cite{Butsky2024} and \cite{Kempski2025a}.

\section{Summary}\label{sec:summary}
To gain a deeper understanding of the transport of highly energetic charged particles, such as cosmic rays, through isotropic structured magnetic turbulence, we studied the motion of test particles in snapshots obtained from a magnetohydrodynamic simulation of a saturated fluctuating dynamo.
Based on a careful analysis of the data, we propose a model that separates particle motion into two distinct modes: non-diffusive magnetised transport along strong coherent flux tubes, and diffusive unmagnetised transport in weak and highly tangled regions of the magnetic field.
We present a stochastic process for each mode, parametrised by separate mean free paths for coherent field line wandering, large-scale mirroring and unmagnetised scattering.
The global transport behaviour emerging from the competition of these processes is described by a combined model, which consists of a stochastic walker that alternates between the two transport modes, where the duration of each segment is sampled from the respective escape-time probability distribution.

The central result of our study is that this combined model accurately reproduces the time-dependent test-particle mean squared displacements, thus providing an explanation for the observed behaviour.
Specifically, field-line wandering and large-scale mirroring with long mean free paths along coherent structures give rise to compound subdiffusion, accounting for the pronounced initial ballistic phase and the transient subdiffusive phase.
Intermittent scattering and unmagnetised motion with short mean free paths facilitate cross-field transport, giving rise to asymptotically diffusive behaviour.

The regime of particle energies, which is currently numerically accessible, is clearly non-asymptotic, i.e. the energy-dependent scaling of the unconditional mean free path does not yet exhibit the expected low-energy power-law behaviour.
However, our results reveal a clear tendency towards the dominance of magnetised motion at small energies,
which implies that field-line wandering, non-resonant mirroring and decoupling of particles from field lines due to intermittent encounters with sharp curvatures are the primary mechanisms for cosmic-ray transport in the asymptotic low-energy regime.
We emphasise that their modelling should be done with care to appropriately account for the highly complex nature of structured turbulence, which is evident in the vastly different length scales characterising individual structures (long coherence lengths and sharp folds), as well as non-trivial correlations between separate structures.

\section*{Acknowledgements}
We gratefully acknowledge helpful discussions with Martin Lemoine, as well as the warm hospitality of Gaetano Zimbardo and Silvia Perri, who organised stimulating workshops at the University of Calabria in September 2024 and February 2025.
J.L. is grateful to Alex Schekochihin and Luca Biferale for the opportunities to present and discuss early stages of this work at the University of Oxford in July 2023 and at the University of Rome Tor Vergata in March 2024, respectively.
J.L., P.R. and S.A. would like to acknowledge the Princeton Center for Theoretical Science for hosting the stimulating workshop ``Synergistic approaches to particle transport in magnetized turbulence: from the laboratory to astrophysics'' in April 2024.
The authors also thank the anonymous referee, whose comments helped to improve this manuscript.

\section*{Funding}
This work was supported by
the Deutsche Forschungsgemeinschaft (DFG, German Research Foundation) through the Collaborative Research Center SFB1491 ``Cosmic Interacting Matters - From Source to Signal'' (grant no.~445052434);
and the International Space Science Institute (ISSI) in Bern through ISSI International Team project \#24-608 ``Energetic Particle Transport in Space Plasma Turbulence''.
The work of P.R. was initially funded through a Gateway Fellowship and subsequently by the DFG through the Walter Benjamin Fellowship (grant no.~518672034).
F.E. acknowledges partial support from NASA LWS grant 80NSSC21K1327.
The authors gratefully acknowledge the Gauss Centre for Supercomputing e.V. (\url{www.gauss-centre.eu}) for funding this project by providing computing time on the SuperMUC-NG at Leibniz Supercomputing Centre (\url{www.lrz.de}) and through the John von Neumann Institute for Computing (NIC) on the GCS Supercomputers JUWELS at Jülich Supercomputing Centre (JSC).

\section*{Declaration of Interests}
The authors report no conflict of interest.

\appendix

\section{Non-dimensionalisation of the relativistic Newton-Lorentz equations}\label{app:lorentz}
The relativistic Newton-Lorentz equations are given by
\begin{equation}
    \dot{\vb{X}\,}\!_t=\vb{V}_t, \quad \dot{\vb{P}\,}\!_t=\frac{q}{m}\left(\vb{E}(\vb{X}_t)+\vb{V}_t\times\vb{B}(\vb{X}_t)\right)
\end{equation}
with the relativistic particle momentum~$\vb{P}_t=\gamma(V_t)\vb{V}_t$.
By writing amplitudes and normalised quantities separately and re-arranging
\begin{align*}
    \frac{V_0}{T_0}\frac{\dd{\hat{\vb{P}\,}\!_t}}{\dd{\hat{t}}}&=\frac{q}{m}V_0B_\mathrm{rms}\hat{\vb{V}}_t\times\hat{\vb{B}\,}\!(\vb{X}_t) + \frac{q}{m}E_\mathrm{rms}\hat{\vb{E}\,}\!(\vb{X}_t), \\
    \frac{\dd{\hat{\vb{P}\,}\!_t}}{\dd{\hat{t}}}&=\alpha\hat{\vb{V}}_t\times\hat{\vb{B}\,}\!(\vb{X}_t) + \beta\hat{\vb{E}\,}\!(\vb{X}_t),
\end{align*}
we can introduce the coupling constants
\begin{equation*}
    \alpha=\frac{q}{m}\frac{L_0B_\mathrm{rms}}{V_0},\quad
    \beta=\frac{q}{m}\frac{L_0E_\mathrm{rms}}{V_0^2},
\end{equation*}
where we replaced the characteristic time~$T_0$ by the outer scale~$L_0=V_0T_0$.
In the limit~$E_\mathrm{rms}\ll V_0$, the particle energy is conserved,~$\gamma(V_t)=\mathrm{const.}$, and we can write
\begin{equation}
    \dot{\vb{V}\,}\!_t=\frac{\dd{\vb{V}_t}}{\dd{t}}=\hat{\omega}_g\vb{V}_t\times\vb{B}(\vb{X}_t),
\end{equation}
where the hats~$\hat{\cdot}$ are dropped for notational convenience and we obtain the normalised gyro frequency
\begin{equation}
    \hat{\omega}_g=\frac{\alpha}{\gamma}=\frac{qB_\mathrm{rms}L_0}{\gamma m v_0}.
\end{equation}

\section{Alternative magnetisation criterion}\label{app:perpcurv}
\begin{figure}
    \centering

	\begin{subfigure}{0.49\linewidth}
        \includegraphics[width=\linewidth]{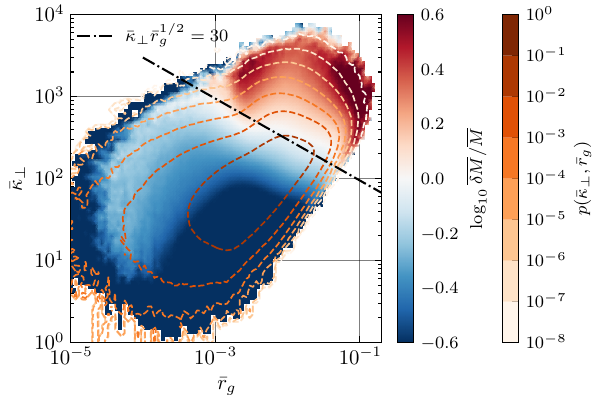}
        \caption{}
		\label{fig:cond-dMperpa}
	\end{subfigure}~%
	\begin{subfigure}{0.49\linewidth}
        \includegraphics[width=\linewidth]{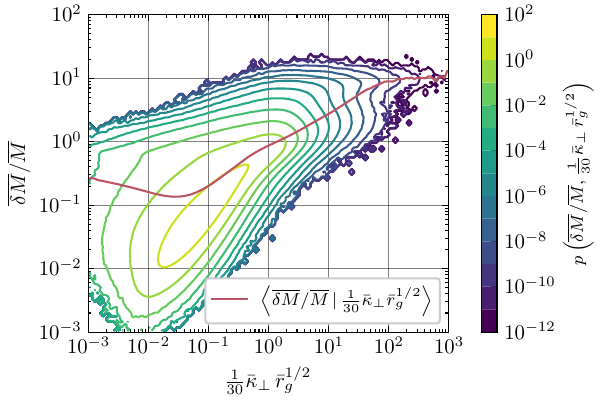}
        \caption{}
		\label{fig:cond-dMperpb}
	\end{subfigure}
    \caption{\textit{(a)}~Average of the relative magnetic moment variation~$\ol{\delta M}/\ol{M}$ conditional on the particle gyro radius~$\bar{r}_g$ and the field-line perpendicular reversal scale~$\kappa_\perp$.
    The transition region~$\ol{\delta M}/\ol{M}\sim1$ between magnetised and unmagnetised transport is approximately described by~$\bar{\kappa}_\perp\bar{r}_g^{1/2}\sim30$.
    The joint density~$p(\bar{\kappa}_\perp,\bar{r}_g)$ is indicated for reference.
    The remarks for the colour scale of figure~\ref{fig:cond-dM} apply here as well.
    \textit{(b)}~Joint density of the relative magnetic moment variation~$\ol{\delta M}/\ol{M}$ and perpendicular magnetisation criterion~$\frac{1}{30}\bar{\kappa}_\perp\bar{r}_g^{1/2}$, as well as the conditional average~$\left\langle\ol{\delta M}/\ol{M}\middle|\frac{1}{30}\bar{\kappa}_\perp\bar{r}_g^{1/2}\right\rangle$.}
    \label{fig:cond-dMperp}
\end{figure}

In addition to the magnetic moment variation conditional on the regular curvature, as given by equation~(\ref{eq:dM}), we also study the dependence on the perpendicular reversal scale \citep{Kempski2023}
\begin{equation}
    \kappa_\perp=\frac{\|(\nabla\times\vb{B})\times\vb{B} - \vb{B}/B\times(\vb{B}\bcdot\nabla\vb{B})\|}{B^2}=\frac{\|\vb{B}\times(\vb{B}\times\nabla\log B)\|}{B^2}.
\end{equation}
The results are shown in figure~\ref{fig:cond-dMperp} and the decision boundary at~$\ol{\delta M}/\ol{M}\sim 1$ is roughly described by
\begin{equation}
    \bar{\kappa}_\perp\bar{r}_g^{1/2}\sim30.
    \label{eq:perp-condition}
\end{equation}

Although this decision boundary appears to be cleaner when compared to figure~\ref{fig:cond-dM}, transport statistics conditional on~$\kappa_\perp$ do not yield significant improvements when compared to~$\kappa$, especially in respect to the bias of the magnetised data.
Additionally, the criterion given by equation~(\ref{eq:perp-condition}) does not convey the compelling physical interpretation of~$\bar{\kappa}\bar{r}_g\sim1$.

\section{Derivations of MSD expressions}\label{app:msd}
\subsection{Langevin equation}\label{app:msd1}
We consider the Langevin equation
\begin{equation}
    \dd{\vb{X}}_s=\vb{V}_s\dd{s},\quad
    \dd{\vb{V}}_s=-\theta\vb{V}_s\dd{s}+\sqrt{\frac{2\theta v^2_\mathrm{rms}}{3}}\dd{\vb{W}}_s,
\end{equation}
with initial conditions~$X_{i,0}=0$ and~$V_{i,0}\sim\mathcal{N}(0, v_\mathrm{rms}^2/3)$ and uncorrelated Gaussian noise~$\langle\dd{W}_{i,s}\dd{W}_{j,s'}\rangle=\delta_{ij}\delta(s-s')$.
The formal solutions of the components of~$\vb{X}_s$ and~$\vb{V}_s$ can be written as
\begin{equation}
    X_{i,s}=\int\limits_{0}^sV_{i,s'}\dd{s'} ,\quad
    V_{i,s}=\sqrt{\frac{2\theta v_\mathrm{rms}^2}{3}}\int\limits_{-\infty}^s e^{-\theta(s-s')} \dd{W}_{s'},
\end{equation}
where the initial conditions are reflected in the lower integral limits.
The auto-covariance function of~$V_{i,s}$ can be computed by means of the Itô isometry as
\begin{align}
    \langle V_{i,r}V_{i,s}\rangle
    &=\frac{2\theta v_\mathrm{rms}^2}{3}e^{-\theta(r+s)} \left\langle\int\limits_{-\infty}^r\int\limits_{-\infty}^s e^{\theta r'} e^{\theta s'} \dd{W}_{r'}\dd{W}_{s'} \right\rangle \nonumber \\
    &=\frac{2\theta v_\mathrm{rms}^2}{3}e^{-\theta(r+s)} \int\limits_{-\infty}^{\min(r,s)}\int\limits_{-\infty}^{\min(r,s)} e^{\theta r'} e^{\theta s'} \dd{r'}\dd{s'} \nonumber \\
    &=\frac{2\theta v_\mathrm{rms}^2}{6\theta} e^{-\theta(r+s)} e^{2\theta\min(r,s)} \nonumber \\
    &=\frac{v_\mathrm{rms}^2}{3}e^{-\theta|r-s|},
\end{align}
which can then be used to find the variance of~$X_{i,s}$ as
\begin{align}
    \langle X_{i,s}^2\rangle
    &=\int\limits_{0}^{s}\int\limits_{0}^{s}\langle V_{i,s'}V_{i,s''}\rangle\dd{s'}\dd{s''} \nonumber \\
    &=\frac{v_\mathrm{rms}^2}{3}\int\limits_{0}^{s}\int\limits_{0}^{s}e^{-\theta|s'-s''|}\dd{s'}\dd{s''} \nonumber \\
    &=\frac{v_\mathrm{rms}^2}{3}\int\limits_{0}^{s}\left(\int\limits_{0}^{s'}e^{-\theta(s'-s'')}\dd{s''}+\int\limits_{s'}^{s}e^{-\theta(s''-s')}\dd{s''}\right)\dd{s'} \nonumber \\
    &=\frac{v_\mathrm{rms}^2}{3}\int\limits_{0}^{s}\left(e^{-\theta s'}\frac{e^{\theta s'}-1}{\theta}-e^{\theta s'}\frac{e^{-\theta s}-e^{-\theta s'}}{\theta}\right)\dd{s'} \nonumber \\
    &=\frac{v_\mathrm{rms}^2}{3\theta}\int\limits_{0}^{s}\left(2-e^{-\theta s'}-e^{-\theta (s-s')}\right)\dd{s'} \nonumber \\
    &=\frac{v_\mathrm{rms}^2}{3\theta}\left(2s+\frac{e^{-\theta s}-1}{\theta}-e^{-\theta s}\frac{e^{\theta s}-1}{\theta}\right) \nonumber \\
    &=\frac{2v_\mathrm{rms}^2}{3\theta^2}\left(e^{-\theta s}+\theta s-1\right).
\end{align}
The variance is equivalent to the MSD, because the processes have zero mean and the initial position is fixed to~$X_{i,0}=0$.

\subsection{Compound subdiffusion}\label{app:msd3}
We find the MSD of the subordinated process
\begin{equation}
    \vb{Y}_t=\vb{X}_{\mathrm{fl},s_t}
\end{equation}
by computing the average MSD of~$\vb{X}_{\mathrm{fl},s}$ weighted by the displacement~$s_t$ at time~$t$
\begin{equation}
    \left\langle Y_{t}^2\right\rangle=\int\limits_{-\infty}^{+\infty}\left\langle X_{\mathrm{fl},|s|}^2\right\rangle \frac{1}{\sqrt{2\upi\langle s_t^2\rangle}} \exp\left(-\frac{s^2}{2\langle s_t^2\rangle}\right)\dd{s},
\end{equation}
where we assume a Gaussian distribution for~$s_t$.
For the short-time asymptotic behaviour, we recall~$\langle X_s^2\rangle\underset{s\to0}{\sim}v_\mathrm{rms}^2s^2$ and~$\langle s_t^2\rangle\underset{t\to0}{\sim}{v_\mathrm{eff}^2t^2}/{3}$, for which we can evaluate the integral as
\begin{align}
\langle Y^2_t\rangle
\underset{t\to0}{\sim}& \int\limits_{-\infty}^{+\infty} v_\mathrm{rms}^2s^2
\frac{1}{\sqrt{2\upi{v_\mathrm{eff}^2t^2}/{3}}}
\exp\left(-\frac{s^2}{2{v_\mathrm{eff}^2t^2}/{3}}\right)\dd{s} \nonumber \\
=& \frac{v_\mathrm{rms}^2}{\sqrt{2\upi}} \frac{\sqrt{3}}{v_\mathrm{eff}t} \int\limits_{-\infty}^{+\infty} s^2
\exp\left(-\frac{3s^2}{2{v_\mathrm{eff}^2t^2}}\right)\dd{s} \nonumber \\
=& \frac{v_\mathrm{rms}^2}{\sqrt{2\upi}} \frac{\sqrt{3}}{v_\mathrm{eff}t} \frac{\sqrt{\upi}}{2\left(\frac{3}{2v_\mathrm{eff}^2t^2}\right)^{3/2}} \nonumber \\
=& \frac{v_\mathrm{rms}^2v_\mathrm{eff}^2 t^2}{3}.
\end{align}

Additionally, for the long-time asymptotic behaviour with~$\langle X_s^2\rangle\underset{s\to\infty}{\sim}{2v_\mathrm{rms}^2}s/{\theta}$ and~$\langle s_t^2\rangle\underset{t\to\infty}{\sim} {2v_\mathrm{eff}^2 t}/{3\theta_\mu}$, we can work out
\begin{align}
\langle Y^2_t) \rangle \underset{t\to\infty}{\sim} & 
\int\limits_{-\infty}^{+\infty} \frac{2v_\mathrm{rms}^2|s|}{\theta} \frac{1}{\sqrt{4\upi {v_\mathrm{eff}^2t}/{3\theta_\mu}}} \exp\left( -\frac{s^2}{4 {v_\mathrm{eff}^2 t}/{3\theta_\mu}} \right) \dd{s} \nonumber \\
= & \frac{v_\mathrm{rms}^2\sqrt{3\theta_\mu}}{v_\mathrm{eff}\theta\sqrt{\upi t}} \int\limits_{-\infty}^{+\infty} |s| \exp\left( -\frac{3\theta_\mu s^2}{4 {v_\mathrm{eff}^2 t}} \right) \dd{s} \nonumber \\
= & \frac{v_\mathrm{rms}^2\sqrt{3\theta_\mu}}{v_\mathrm{eff}\theta\sqrt{\upi t}} \sqrt{\upi} \frac{4 {v_\mathrm{eff}^2 t}}{3\theta_\mu} \nonumber \\
= & \frac{4 {v_\mathrm{rms}^2 v_\mathrm{eff}\sqrt{t}}}{\theta\sqrt{3\theta_\mu}} \nonumber \\
= & \frac{4v_\mathrm{rms} \lambda_\mathrm{fl}\sqrt{v_\mathrm{eff}\lambda_\mu t}}{3},
\end{align}
where we made use of the MFP expressions~$\lambda_\mathrm{fl}=v_\mathrm{rms}/\theta$ and~$\lambda_\mu=v_\mathrm{eff}/\theta_\mu$.

\section{Bayesian optimisation}\label{app:bayes}
Optimising the loss function given by equation~(\ref{eq:loss}) requires
the simulation of a sufficiently large number of samples of our combined stochastic model given by Algorithm~\ref{alg:combined}
at each step of the optimisation procedure.
The number of samples needs to be large enough to produce a sufficiently converged average~$\langle\Delta Z_{\mathrm{model},\tau}^2\rangle$, but also small enough to remain within reasonable computational cost.
However, even for a carefully chosen number of samples, evaluating equation~(\ref{eq:loss}) remains relatively expensive and noisy.

\begin{figure}
    \centering
    \includegraphics[width=\linewidth]{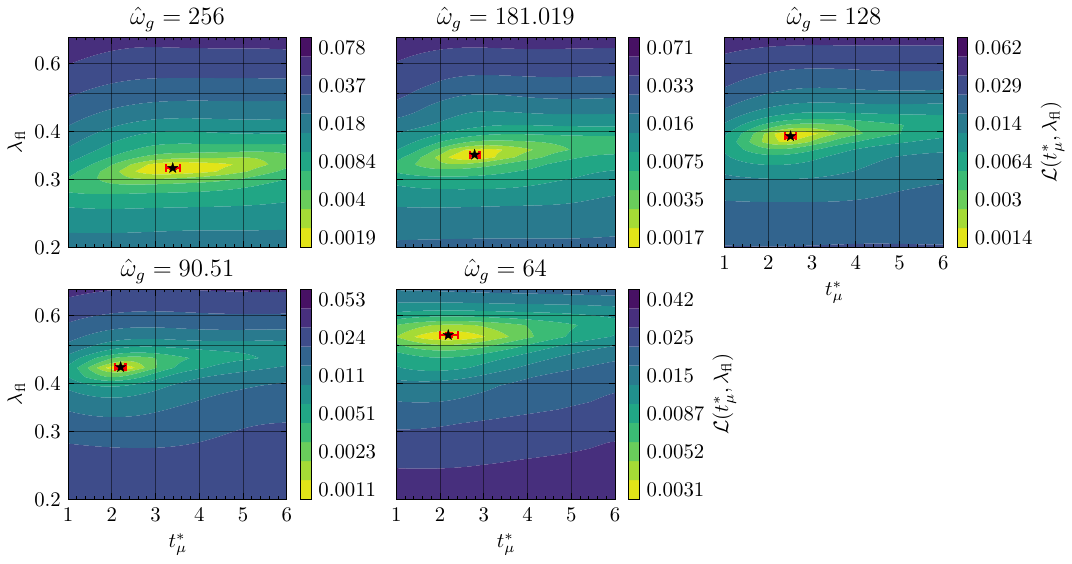}
    \caption{Landscapes of the loss function equation~(\ref{eq:loss}) as estimated by Bayesian optimisation, including the expected minimum and estimated confidence intervals.}
    \label{fig:loss}
\end{figure}

We therefore turn to Bayesian optimisation as implemented in the software package \texttt{scikit-optimize} \citep{Head2021}, which treats evaluations of the loss function as samples drawn from a Gaussian process.
Based on previous observations, a cheap acquisition function is optimised, which gives the next point in the parameter space to evaluate.
In this way, a reliable estimate of the optimal parameters is achievable for a small number of evaluations of the loss function.
The procedure additionally provides a hyper-parameter~$\xi$ to balance exploration of uncertain regions of the parameter space against exploitation of regions, which likely contain a minimum of the loss function.

For each~$\hat{\omega}_g$, we run 64 steps with~$\xi=1$ for exploration, followed by 64 steps with~$\xi=0.01$ for exploitation of likely minima.
For $96\,000$ samples of the combined stochastic model, distributed on 96 CPU cores, a single evaluation of the loss function takes approximately 35 seconds.
The resulting loss functions are shown in figure~\ref{fig:loss}, including the expected minima and roughly estimated confidence intervals.

Confidence intervals for Bayesian optimisation are not provided by the software package and hard to find in the literature.
We thus produce, for each $\hat{\omega}_g$, a naive estimate based on 480 independent realisations of the underlying trained Gaussian process.
The realisations are generated for random points on the parameter grid in the neighbourhood of the expected minimum and the minimum of each realisation is recorded.
The mean of the minima of the realisations converges to the expected minimum and their standard deviation is taken as confidence bounds on the parameters.

\bibliographystyle{jpp}
\bibliography{main}

\begin{thebibliography}{138}
\expandafter\ifx\csname natexlab\endcsname\relax\def\natexlab#1{#1}\fi
\def\au#1{#1} \def\ed#1{#1} \def\yr#1{#1}\def\at#1{#1}\def\jt#1{\textit{#1}} \def\bt#1{#1}\def\bvol#1{\textbf{#1}} \def\vol#1{#1} \def\pg#1{#1} \def\publ#1{#1}\def\arxiv#1{#1}\def\org#1{#1}\def\st#1{\textit{#1}}

\bibitem[{Aerdker} {\em et~al.\/}(2025){Aerdker}, {Merten}, {Effenberger}, {Fichtner} \& {Becker Tjus}]{Aerdker2025}
{\sc \au{{Aerdker}, Sophie}, \au{{Merten}, Lukas}, \au{{Effenberger}, Frederic}, \au{{Fichtner}, Horst} \& \au{{Becker Tjus}, Julia}} \yr{2025}  \at{{Superdiffusion of energetic particles at shocks: A L{\'e}vy flight model for acceleration}}.  \jt{\aap}  \bvol{693},  \pg{A15}.

\bibitem[{Albright} {\em et~al.\/}(2001){Albright}, {Chandran}, {Cowley} \& {Loh}]{Albright2001}
{\sc \au{{Albright}, B.~J.}, \au{{Chandran}, B.~D.~G.}, \au{{Cowley}, S.~C.} \& \au{{Loh}, M.}} \yr{2001}  \at{{Parallel heat diffusion and subdiffusion in random magnetic fields}}.  \jt{Physics of Plasmas}  \bvol{8}~(3),  \pg{777--787}.

\bibitem[{Alvelius}(1999)]{Alvelius1999}
{\sc \au{{Alvelius}, K.}} \yr{1999}  \at{{Random forcing of three-dimensional homogeneous turbulence}}.  \jt{Physics of Fluids}  \bvol{11}~(7),  \pg{1880--1889}.

\bibitem[{Amato} \& {Blasi}(2018)]{Amato2018}
{\sc \au{{Amato}, Elena} \& \au{{Blasi}, Pasquale}} \yr{2018}  \at{{Cosmic ray transport in the Galaxy: A review}}.  \jt{Advances in Space Research}  \bvol{62}~(10),  \pg{2731--2749}.

\bibitem[{Armillotta} {\em et~al.\/}(2022){Armillotta}, {Ostriker} \& {Jiang}]{Armillotta2022}
{\sc \au{{Armillotta}, Lucia}, \au{{Ostriker}, Eve~C.} \& \au{{Jiang}, Yan-Fei}} \yr{2022}  \at{{Cosmic-Ray Transport in Varying Galactic Environments}}.  \jt{\apj}  \bvol{929}~(2),  \pg{170}.

\bibitem[{Arzner} {\em et~al.\/}(2006){Arzner}, {Knaepen}, {Carati}, {Denewet} \& {Vlahos}]{Arzner2006}
{\sc \au{{Arzner}, Kaspar}, \au{{Knaepen}, Bernard}, \au{{Carati}, Daniele}, \au{{Denewet}, Nicolas} \& \au{{Vlahos}, Loukas}} \yr{2006}  \at{{The Effect of Coherent Structures on Stochastic Acceleration in MHD Turbulence}}.  \jt{\apj}  \bvol{637}~(1),  \pg{322--332}.

\bibitem[{Baggaley} {\em et~al.\/}(2010){Baggaley}, {Shukurov}, {Barenghi} \& {Subramanian}]{Baggaley2010}
{\sc \au{{Baggaley}, A.~W.}, \au{{Shukurov}, A.}, \au{{Barenghi}, C.~F.} \& \au{{Subramanian}, K.}} \yr{2010}  \at{{Fluctuation dynamo based on magnetic reconnections}}.  \jt{Astronomische Nachrichten}  \bvol{331}~(1),  \pg{46}.

\bibitem[{Bai} {\em et~al.\/}(2015){Bai}, {Caprioli}, {Sironi} \& {Spitkovsky}]{Bai2015}
{\sc \au{{Bai}, Xue-Ning}, \au{{Caprioli}, Damiano}, \au{{Sironi}, Lorenzo} \& \au{{Spitkovsky}, Anatoly}} \yr{2015}  \at{{Magnetohydrodynamic-particle-in-cell Method for Coupling Cosmic Rays with a Thermal Plasma: Application to Non-relativistic Shocks}}.  \jt{\apj}  \bvol{809}~(1),  \pg{55}.

\bibitem[{Balcerek} {\em et~al.\/}(2023){Balcerek}, {Wy{\l}oma{\'n}ska}, {Burnecki}, {Metzler} \& {Krapf}]{Balcerek2023}
{\sc \au{{Balcerek}, Micha{\l}}, \au{{Wy{\l}oma{\'n}ska}, Agnieszka}, \au{{Burnecki}, Krzysztof}, \au{{Metzler}, Ralf} \& \au{{Krapf}, Diego}} \yr{2023}  \at{{Modelling intermittent anomalous diffusion with switching fractional Brownian motion}}.  \jt{New Journal of Physics}  \bvol{25}~(10),  \pg{103031}.

\bibitem[{Balescu} {\em et~al.\/}(1994){Balescu}, {Wang} \& {Misguich}]{Balescu1994}
{\sc \au{{Balescu}, R.}, \au{{Wang}, Hai-Da} \& \au{{Misguich}, J.~H.}} \yr{1994}  \at{{Langevin equation versus kinetic equation: Subdiffusive behavior of charged particles in a stochastic magnetic field}}.  \jt{Physics of Plasmas}  \bvol{1}~(12),  \pg{3826--3842}.

\bibitem[{Bandyopadhyay} {\em et~al.\/}(2020){Bandyopadhyay}, {Yang}, {Matthaeus}, {Chasapis}, {Parashar}, {Russell}, {Strangeway}, {Torbert}, {Giles}, {Gershman}, {Pollock}, {Moore} \& {Burch}]{Bandyopadhyay2020}
{\sc \au{{Bandyopadhyay}, Riddhi}, \au{{Yang}, Yan}, \au{{Matthaeus}, William~H.}, \au{{Chasapis}, Alexandros}, \au{{Parashar}, Tulasi~N.}, \au{{Russell}, Christopher~T.}, \au{{Strangeway}, Robert~J.}, \au{{Torbert}, Roy~B.}, \au{{Giles}, Barbara~L.}, \au{{Gershman}, Daniel~J.}, \au{{Pollock}, Craig~J.}, \au{{Moore}, Thomas~E.} \& \au{{Burch}, James~L.}} \yr{2020}  \at{{In Situ Measurement of Curvature of Magnetic Field in Turbulent Space Plasmas: A Statistical Study}}.  \jt{\apjl}  \bvol{893}~(1),  \pg{L25}.

\bibitem[{Beattie} {\em et~al.\/}(2024){Beattie}, {Federrath}, {Klessen}, {Cielo} \& {Bhattacharjee}]{Beattie2024}
{\sc \au{{Beattie}, James~R.}, \au{{Federrath}, Christoph}, \au{{Klessen}, Ralf~S.}, \au{{Cielo}, Salvatore} \& \au{{Bhattacharjee}, Amitava}} \yr{2024}  \at{{Magnetized compressible turbulence with a fluctuation dynamo and Reynolds numbers over a million}}.  \jt{arXiv e-prints}  \pg{p. arXiv:2405.16626}.

\bibitem[{Beattie} {\em et~al.\/}(2025){Beattie}, {Noer Kolborg}, {Ramirez-Ruiz} \& {Federrath}]{Beattie2025b}
{\sc \au{{Beattie}, James~R.}, \au{{Noer Kolborg}, Anne}, \au{{Ramirez-Ruiz}, Enrico} \& \au{{Federrath}, Christoph}} \yr{2025}  \at{{So long Kolmogorov: the forward and backward turbulence cascades in a supernovae-driven, multiphase interstellar medium}}.  \jt{arXiv e-prints}  \pg{p. arXiv:2501.09855}.

\bibitem[{Bell} {\em et~al.\/}(2025){Bell}, {Matthews}, {Taylor} \& {Giacinti}]{Bell2025}
{\sc \au{{Bell}, A.~R.}, \au{{Matthews}, J.~H.}, \au{{Taylor}, A.~M.} \& \au{{Giacinti}, G.}} \yr{2025}  \at{{Cosmic ray transport and acceleration with magnetic mirroring}}.  \jt{\mnras}  \bvol{539}~(2),  \pg{1236--1247}.

\bibitem[{Beresnyak} {\em et~al.\/}(2011){Beresnyak}, {Yan} \& {Lazarian}]{Beresnyak2011}
{\sc \au{{Beresnyak}, A.}, \au{{Yan}, H.} \& \au{{Lazarian}, A.}} \yr{2011}  \at{{Numerical Study of Cosmic Ray Diffusion in Magnetohydrodynamic Turbulence}}.  \jt{\apj}  \bvol{728}~(1),  \pg{60}.

\bibitem[{Bian} \& {Li}(2024)]{Bian2024}
{\sc \au{{Bian}, N.~H.} \& \au{{Li}, Gang}} \yr{2024}  \at{{Lagrangian Perspectives on the Small-scale Structure of Alfv{\'e}nic Turbulence and Stochastic Models for the Dispersion of Fluid Particles and Magnetic Field Lines in the Solar Wind}}.  \jt{\apjs}  \bvol{273}~(1),  \pg{15}.

\bibitem[{Biskamp}(2003)]{Biskamp2003}
{\sc \au{{Biskamp}, Dieter}} \yr{2003} {\em {Magnetohydrodynamic Turbulence}\/}.  \publ{Cambridge, UK: Cambridge University Press}.

\bibitem[{Boldyrev}(2006)]{Boldyrev2006}
{\sc \au{{Boldyrev}, Stanislav}} \yr{2006}  \at{{Spectrum of Magnetohydrodynamic Turbulence}}.  \jt{\prl}  \bvol{96}~(11),  \pg{115002}.

\bibitem[Boris \& Shanny(1971)]{boris:1971}
{\sc \au{Boris, Jay~P} \& \au{Shanny, Ramy~A}} \yr{1971} {\em Proceedings, Fourth Conference on Numerical Simulation of Plasmas\/}.  \publ{Naval Research Laboratory}.

\bibitem[{B{\"o}ss} {\em et~al.\/}(2023){B{\"o}ss}, {Steinwandel}, {Dolag} \& {Lesch}]{Boss2023}
{\sc \au{{B{\"o}ss}, Ludwig~M.}, \au{{Steinwandel}, Ulrich~P.}, \au{{Dolag}, Klaus} \& \au{{Lesch}, Harald}} \yr{2023}  \at{{CRESCENDO: an on-the-fly Fokker-Planck solver for spectral cosmic rays in cosmological simulations}}.  \jt{\mnras}  \bvol{519}~(1),  \pg{548--572}.

\bibitem[{Branch} {\em et~al.\/}(1999){Branch}, {Coleman} \& {Li}]{Branch1999}
{\sc \au{{Branch}, Mary~Ann}, \au{{Coleman}, Thomas~F.} \& \au{{Li}, Yuying}} \yr{1999}  \at{{A Subspace, Interior, and Conjugate Gradient Method for Large-Scale Bound-Constrained Minimization Problems}}.  \jt{SIAM Journal on Scientific Computing}  \bvol{21}~(1),  \pg{1--23}.

\bibitem[{Brandenburg} \& {Sarson}(2002)]{Brandenburg2002}
{\sc \au{{Brandenburg}, Axel} \& \au{{Sarson}, Graeme~R.}} \yr{2002}  \at{{Effect of Hyperdiffusivity on Turbulent Dynamos with Helicity}}.  \jt{\prl}  \bvol{88}~(5),  \pg{055003}.

\bibitem[{Bressloff} \& {Lawley}(2017)]{Bressloff2017}
{\sc \au{{Bressloff}, Paul~C.} \& \au{{Lawley}, Sean~D.}} \yr{2017}  \at{{Temporal disorder as a mechanism for spatially heterogeneous diffusion}}.  \jt{\pre}  \bvol{95}~(6),  \pg{060101}.

\bibitem[Burekovi{\'c} {\em et~al.\/}(2024)Burekovi{\'c}, Sch{\"a}fer \& Grauer]{burekovic-schaefer-grauer:2024}
{\sc \au{Burekovi{\'c}, Sumeja}, \au{Sch{\"a}fer, Tobias} \& \au{Grauer, Rainer}} \yr{2024}  \at{Instantons, fluctuations, and singularities in the supercritical stochastic nonlinear schr{\"o}dinger equation}.  \jt{Phys. Rev. Lett.}  \bvol{133}~(7),  \pg{077202}.

\bibitem[{Butsky} {\em et~al.\/}(2024){Butsky}, {Hopkins}, {Kempski}, {Ponnada}, {Quataert} \& {Squire}]{Butsky2024}
{\sc \au{{Butsky}, Iryna~S.}, \au{{Hopkins}, Philip~F.}, \au{{Kempski}, Philipp}, \au{{Ponnada}, Sam~B.}, \au{{Quataert}, Eliot} \& \au{{Squire}, Jonathan}} \yr{2024}  \at{{Galactic cosmic-ray scattering due to intermittent structures}}.  \jt{\mnras}  \bvol{528}~(3),  \pg{4245--4254}.

\bibitem[{Chandran} {\em et~al.\/}(1999){Chandran}, {Cowley}, {Ivanushkina} \& {Sydora}]{Chandran1999}
{\sc \au{{Chandran}, Benjamin D.~G.}, \au{{Cowley}, Steven~C.}, \au{{Ivanushkina}, Mariya} \& \au{{Sydora}, Richard}} \yr{1999}  \at{{Heat Transport Along an Inhomogeneous Magnetic Field. I. Periodic Magnetic Mirrors}}.  \jt{\apj}  \bvol{525}~(2),  \pg{638--650}.

\bibitem[{Chandran} {\em et~al.\/}(2015){Chandran}, {Schekochihin} \& {Mallet}]{Chandran2015}
{\sc \au{{Chandran}, B.~D.~G.}, \au{{Schekochihin}, A.~A.} \& \au{{Mallet}, A.}} \yr{2015}  \at{{Intermittency and Alignment in Strong RMHD Turbulence}}.  \jt{\apj}  \bvol{807}~(1),  \pg{39}.

\bibitem[{Chandrasekhar}(1943)]{Chandrasekhar1943}
{\sc \au{{Chandrasekhar}, S.}} \yr{1943}  \at{{Stochastic Problems in Physics and Astronomy}}.  \jt{Reviews of Modern Physics}  \bvol{15}~(1),  \pg{1--89}.

\bibitem[{Chechkin} {\em et~al.\/}(2017){Chechkin}, {Seno}, {Metzler} \& {Sokolov}]{Chechkin2017}
{\sc \au{{Chechkin}, Aleksei~V.}, \au{{Seno}, Flavio}, \au{{Metzler}, Ralf} \& \au{{Sokolov}, Igor~M.}} \yr{2017}  \at{{Brownian yet Non-Gaussian Diffusion: From Superstatistics to Subordination of Diffusing Diffusivities}}.  \jt{Physical Review X}  \bvol{7}~(2),  \pg{021002}.

\bibitem[{Chen} {\em et~al.\/}(2020){Chen}, {Bale}, {Bonnell}, {Borovikov}, {Bowen}, {Burgess}, {Case}, {Chandran}, {de Wit}, {Goetz}, {Harvey}, {Kasper}, {Klein}, {Korreck}, {Larson}, {Livi}, {MacDowall}, {Malaspina}, {Mallet}, {McManus}, {Moncuquet}, {Pulupa}, {Stevens} \& {Whittlesey}]{Chen2020b}
{\sc \au{{Chen}, C.~H.~K.}, \au{{Bale}, S.~D.}, \au{{Bonnell}, J.~W.}, \au{{Borovikov}, D.}, \au{{Bowen}, T.~A.}, \au{{Burgess}, D.}, \au{{Case}, A.~W.}, \au{{Chandran}, B.~D.~G.}, \au{{de Wit}, T.~Dudok}, \au{{Goetz}, K.}, \au{{Harvey}, P.~R.}, \au{{Kasper}, J.~C.}, \au{{Klein}, K.~G.}, \au{{Korreck}, K.~E.}, \au{{Larson}, D.}, \au{{Livi}, R.}, \au{{MacDowall}, R.~J.}, \au{{Malaspina}, D.~M.}, \au{{Mallet}, A.}, \au{{McManus}, M.~D.}, \au{{Moncuquet}, M.}, \au{{Pulupa}, M.}, \au{{Stevens}, M.~L.} \& \au{{Whittlesey}, P.}} \yr{2020}  \at{{The Evolution and Role of Solar Wind Turbulence in the Inner Heliosphere}}.  \jt{\apjs}  \bvol{246}~(2),  \pg{53}.

\bibitem[{Chernoglazov} {\em et~al.\/}(2021){Chernoglazov}, {Ripperda} \& {Philippov}]{Chernoglazov2021}
{\sc \au{{Chernoglazov}, Alexander}, \au{{Ripperda}, Bart} \& \au{{Philippov}, Alexander}} \yr{2021}  \at{{Dynamic Alignment and Plasmoid Formation in Relativistic Magnetohydrodynamic Turbulence}}.  \jt{\apjl}  \bvol{923}~(1),  \pg{L13}.

\bibitem[{Cohet} \& {Marcowith}(2016)]{Cohet2016}
{\sc \au{{Cohet}, R.} \& \au{{Marcowith}, A.}} \yr{2016}  \at{{Cosmic ray propagation in sub-Alfv{\'e}nic magnetohydrodynamic turbulence}}.  \jt{\aap}  \bvol{588},  \pg{A73}.

\bibitem[{Comisso} \& {Sironi}(2019)]{Comisso2019}
{\sc \au{{Comisso}, Luca} \& \au{{Sironi}, Lorenzo}} \yr{2019}  \at{{The Interplay of Magnetically Dominated Turbulence and Magnetic Reconnection in Producing Nonthermal Particles}}.  \jt{\apj}  \bvol{886}~(2),  \pg{122}.

\bibitem[{Crutcher}(2012)]{Crutcher-2012}
{\sc \au{{Crutcher}, Richard~M.}} \yr{2012}  \at{{Magnetic Fields in Molecular Clouds}}.  \jt{\araa}  \bvol{50},  \pg{29--63}.

\bibitem[{Datta} {\em et~al.\/}(2024){Datta}, {Beta} \& {Gro{\ss}mann}]{Datta2024}
{\sc \au{{Datta}, Agniva}, \au{{Beta}, Carsten} \& \au{{Gro{\ss}mann}, Robert}} \yr{2024}  \at{{Random walks of intermittently self-propelled particles}}.  \jt{Physical Review Research}  \bvol{6}~(4),  \pg{043281}.

\bibitem[{Doerries} {\em et~al.\/}(2022){Doerries}, {Chechkin} \& {Metzler}]{Doerries2022}
{\sc \au{{Doerries}, T.}, \au{{Chechkin}, A.~V.} \& \au{{Metzler}, R.}} \yr{2022}  \at{{Apparent anomalous diffusion and non-Gaussian distributions in a simple mobile-immobile transport model with Poissonian switching}}.  \jt{arXiv e-prints}  \pg{p. arXiv:2203.13328}.

\bibitem[{Dom{\'\i}nguez Fern{\'a}ndez}(2020)]{DominguezFernandez-2020}
{\sc \au{{Dom{\'\i}nguez Fern{\'a}ndez}, Paola}} \yr{2020}  \at{{Magnetic fields in the intracluster medium}}. PhD thesis, University of Hamburg, Germany.

\bibitem[{Dong} {\em et~al.\/}(2022){Dong}, {Wang}, {Huang}, {Comisso}, {Sandstrom} \& {Bhattacharjee}]{Dong2022}
{\sc \au{{Dong}, Chuanfei}, \au{{Wang}, Liang}, \au{{Huang}, Yi-Min}, \au{{Comisso}, Luca}, \au{{Sandstrom}, Timothy~A.} \& \au{{Bhattacharjee}, Amitava}} \yr{2022}  \at{{Reconnection-driven energy cascade in magnetohydrodynamic turbulence}}.  \jt{Science Advances}  \bvol{8}~(49),  \pg{eabn7627}.

\bibitem[{D{\"o}rner} {\em et~al.\/}(2023){D{\"o}rner}, {Reichherzer}, {Becker Tjus} \& {Heesen}]{Dorner2023}
{\sc \au{{D{\"o}rner}, Julien}, \au{{Reichherzer}, Patrick}, \au{{Becker Tjus}, Julia} \& \au{{Heesen}, Volker}} \yr{2023}  \at{{Cosmic-ray electron transport in the galaxy M 51}}.  \jt{\aap}  \bvol{669},  \pg{A111}.

\bibitem[Drummond(1982)]{drummond:1982}
{\sc \au{Drummond, I.~T.}} \yr{1982}  \at{Path-integral methods for turbulent diffusion}.  \jt{Journal of Fluid Mechanics}  \bvol{123},  \pg{59–68}.

\bibitem[{Durrive} {\em et~al.\/}(2022){Durrive}, {Changmai}, {Keppens}, {Lesaffre}, {Maci} \& {Momferatos}]{Durrive2022}
{\sc \au{{Durrive}, Jean-Baptiste}, \au{{Changmai}, Madhurjya}, \au{{Keppens}, Rony}, \au{{Lesaffre}, Pierre}, \au{{Maci}, Daniela} \& \au{{Momferatos}, Georgios}} \yr{2022}  \at{{Swift generator for three-dimensional magnetohydrodynamic turbulence}}.  \jt{\pre}  \bvol{106}~(2),  \pg{025307}.

\bibitem[{Effenberger} {\em et~al.\/}(2024){Effenberger}, {Aerdker}, {Merten} \& {Fichtner}]{Effenberger2024}
{\sc \au{{Effenberger}, Frederic}, \au{{Aerdker}, Sophie}, \au{{Merten}, Lukas} \& \au{{Fichtner}, Horst}} \yr{2024}  \at{{Superdiffusion of energetic particles at shocks: A fractional diffusion and L{\'e}vy flight model of spatial transport}}.  \jt{\aap}  \bvol{686},  \pg{A219}.

\bibitem[{Elmegreen} \& {Scalo}(2004)]{Elmegreen2004}
{\sc \au{{Elmegreen}, Bruce~G.} \& \au{{Scalo}, John}} \yr{2004}  \at{{Interstellar Turbulence I: Observations and Processes}}.  \jt{\araa}  \bvol{42}~(1),  \pg{211--273}.

\bibitem[{Els} {\em et~al.\/}(2024){Els}, {Engelbrecht}, {Lang} \& {Strauss}]{Els2024a}
{\sc \au{{Els}, P.~L.}, \au{{Engelbrecht}, N.~E.}, \au{{Lang}, J.~T.} \& \au{{Strauss}, R.~D.}} \yr{2024}  \at{{The Diffusion Tensor of Protons at 1 au: Comparing Simulation, Observation, and Theory}}.  \jt{\apj}  \bvol{975}~(1),  \pg{134}.

\bibitem[{Engelbrecht} {\em et~al.\/}(2022){Engelbrecht}, {Effenberger}, {Florinski}, {Potgieter}, {Ruffolo}, {Chhiber}, {Usmanov}, {Rankin} \& {Els}]{Engelbrecht2022}
{\sc \au{{Engelbrecht}, N.~Eugene}, \au{{Effenberger}, F.}, \au{{Florinski}, V.}, \au{{Potgieter}, M.~S.}, \au{{Ruffolo}, D.}, \au{{Chhiber}, R.}, \au{{Usmanov}, A.~V.}, \au{{Rankin}, J.~S.} \& \au{{Els}, P.~L.}} \yr{2022}  \at{{Theory of Cosmic Ray Transport in the Heliosphere}}.  \jt{\ssr}  \bvol{218}~(4),  \pg{33}.

\bibitem[{Ewart} {\em et~al.\/}(2024){Ewart}, {Reichherzer}, {Bott}, {Kunz} \& {Schekochihin}]{Ewart2024}
{\sc \au{{Ewart}, Robert~J.}, \au{{Reichherzer}, Patrick}, \au{{Bott}, Archie F.~A.}, \au{{Kunz}, Matthew~W.} \& \au{{Schekochihin}, Alexander~A.}} \yr{2024}  \at{{Cosmic-ray confinement in radio bubbles by micromirrors}}.  \jt{\mnras}  \bvol{532}~(2),  \pg{2098--2107}.

\bibitem[{Galishnikova} {\em et~al.\/}(2022){Galishnikova}, {Kunz} \& {Schekochihin}]{Galishnikova2022}
{\sc \au{{Galishnikova}, Alisa~K.}, \au{{Kunz}, Matthew~W.} \& \au{{Schekochihin}, Alexander~A.}} \yr{2022}  \at{{Tearing Instability and Current-Sheet Disruption in the Turbulent Dynamo}}.  \jt{Physical Review X}  \bvol{12}~(4),  \pg{041027}.

\bibitem[{Gent} {\em et~al.\/}(2024){Gent}, {Mac Low} \& {Korpi-Lagg}]{Gent2024}
{\sc \au{{Gent}, Frederick~A.}, \au{{Mac Low}, Mordecai-Mark} \& \au{{Korpi-Lagg}, Maarit~J.}} \yr{2024}  \at{{Transition from Small-scale to Large-scale Dynamo in a Supernova-driven, Multiphase Medium}}.  \jt{\apj}  \bvol{961}~(1),  \pg{7}.

\bibitem[Grafke {\em et~al.\/}(2015)Grafke, Grauer \& Sch\"afer]{grafke-grauer-schaefer:2015c}
{\sc \au{Grafke, Tobias}, \au{Grauer, Rainer} \& \au{Sch\"afer, Tobias}} \yr{2015}  \at{The instanton method and its numerical implementation in fluid mechanics}.  \jt{J. Phys. A Math. Theor.}  \bvol{48}~(33),  \pg{333001}.

\bibitem[{Grauer} {\em et~al.\/}(1994){Grauer}, {Krug} \& {Marliani}]{Grauer1994}
{\sc \au{{Grauer}, R.}, \au{{Krug}, J.} \& \au{{Marliani}, C.}} \yr{1994}  \at{{Scaling of high-order structure functions in magnetohydrodynamic turbulence}}.  \jt{Physics Letters A}  \bvol{195}~(5-6),  \pg{335--338}.

\bibitem[{Grauer} \& {Marliani}(2000)]{Grauer2000}
{\sc \au{{Grauer}, Rainer} \& \au{{Marliani}, Christiane}} \yr{2000}  \at{{Current-Sheet Formation in 3D Ideal Incompressible Magnetohydrodynamics}}.  \jt{\prl}  \bvol{84}~(21),  \pg{4850--4853}.

\bibitem[{Gro{\v{s}}elj} {\em et~al.\/}(2019){Gro{\v{s}}elj}, {Chen}, {Mallet}, {Samtaney}, {Schneider} \& {Jenko}]{Groselj2019}
{\sc \au{{Gro{\v{s}}elj}, Daniel}, \au{{Chen}, Christopher H.~K.}, \au{{Mallet}, Alfred}, \au{{Samtaney}, Ravi}, \au{{Schneider}, Kai} \& \au{{Jenko}, Frank}} \yr{2019}  \at{{Kinetic Turbulence in Astrophysical Plasmas: Waves and/or Structures?}}  \jt{Physical Review X}  \bvol{9}~(3),  \pg{031037}.

\bibitem[{Haugen} \& {Brandenburg}(2004)]{Haugen2004b}
{\sc \au{{Haugen}, Nils Erland~L.} \& \au{{Brandenburg}, Axel}} \yr{2004}  \at{{Inertial range scaling in numerical turbulence with hyperviscosity}}.  \jt{\pre}  \bvol{70}~(2),  \pg{026405}.

\bibitem[{Head} {\em et~al.\/}(2021){Head}, {Kumar}, {Nahrstaedt}, {Louppe} \& {Shcherbatyi}]{Head2021}
{\sc \au{{Head}, Tim}, \au{{Kumar}, Manoj}, \au{{Nahrstaedt}, Holger}, \au{{Louppe}, Gilles} \& \au{{Shcherbatyi}, Iaroslav}} \yr{2021} {scikit-optimize/scikit-optimize}.

\bibitem[{Hopkins} {\em et~al.\/}(2020){Hopkins}, {Chan}, {Garrison-Kimmel}, {Ji}, {Su}, {Hummels}, {Kere{\v{s}}}, {Quataert} \& {Faucher-Gigu{\`e}re}]{Hopkins2020}
{\sc \au{{Hopkins}, Philip~F.}, \au{{Chan}, T.~K.}, \au{{Garrison-Kimmel}, Shea}, \au{{Ji}, Suoqing}, \au{{Su}, Kung-Yi}, \au{{Hummels}, Cameron~B.}, \au{{Kere{\v{s}}}, Du{\v{s}}an}, \au{{Quataert}, Eliot} \& \au{{Faucher-Gigu{\`e}re}, Claude-Andr{\'e}}} \yr{2020}  \at{{But what about...: cosmic rays, magnetic fields, conduction, and viscosity in galaxy formation}}.  \jt{\mnras}  \bvol{492}~(3),  \pg{3465--3498}.

\bibitem[{Hopkins} {\em et~al.\/}(2022){Hopkins}, {Squire}, {Butsky} \& {Ji}]{Hopkins2022a}
{\sc \au{{Hopkins}, Philip~F.}, \au{{Squire}, Jonathan}, \au{{Butsky}, Iryna~S.} \& \au{{Ji}, Suoqing}} \yr{2022}  \at{{Standard self-confinement and extrinsic turbulence models for cosmic ray transport are fundamentally incompatible with observations}}.  \jt{\mnras}  \bvol{517}~(4),  \pg{5413--5448}.

\bibitem[{Hosking} \& {Schekochihin}(2021)]{Hosking2021}
{\sc \au{{Hosking}, David~N.} \& \au{{Schekochihin}, Alexander~A.}} \yr{2021}  \at{{Reconnection-Controlled Decay of Magnetohydrodynamic Turbulence and the Role of Invariants}}.  \jt{Physical Review X}  \bvol{11}~(4),  \pg{041005}.

\bibitem[{Houde} {\em et~al.\/}(2009){Houde}, {Vaillancourt}, {Hildebrand}, {Chitsazzadeh} \& {Kirby}]{Houde-etal-2009}
{\sc \au{{Houde}, Martin}, \au{{Vaillancourt}, John~E.}, \au{{Hildebrand}, Roger~H.}, \au{{Chitsazzadeh}, Shadi} \& \au{{Kirby}, Larry}} \yr{2009}  \at{{Dispersion of Magnetic Fields in Molecular Clouds. II.}}  \jt{\apj}  \bvol{706}~(2),  \pg{1504--1516}.

\bibitem[Hudson \& Li(2020)]{hudson-li:2020}
{\sc \au{Hudson, Thomas} \& \au{Li, Xingjie~H.}} \yr{2020}  \at{Coarse-graining of overdamped langevin dynamics via the mori--zwanzig formalism}.  \jt{Multiscale Modeling \& Simulation}  \bvol{18}~(2),  \pg{1113--1135}.

\bibitem[{Jansson} \& {Farrar}(2012)]{Jansson-Farrar-2012}
{\sc \au{{Jansson}, Ronnie} \& \au{{Farrar}, Glennys~R.}} \yr{2012}  \at{{A New Model of the Galactic Magnetic Field}}.  \jt{\apj}  \bvol{757}~(1),  \pg{14}.

\bibitem[{Jokipii}(1966)]{Jokipii1966}
{\sc \au{{Jokipii}, J.~R.}} \yr{1966}  \at{{Cosmic-Ray Propagation. I. Charged Particles in a Random Magnetic Field}}.  \jt{\apj}  \bvol{146},  \pg{480}.

\bibitem[{Kamal Youssef} \& {Grenier}(2024)]{KamalYoussef2024}
{\sc \au{{Kamal Youssef}, F.~R.} \& \au{{Grenier}, I.~A.}} \yr{2024}  \at{{Cosmic-ray diffusion in two local filamentary clouds}}.  \jt{\aap}  \bvol{685},  \pg{A102}.

\bibitem[{Kempski} {\em et~al.\/}(2023){Kempski}, {Fielding}, {Quataert}, {Galishnikova}, {Kunz}, {Philippov} \& {Ripperda}]{Kempski2023}
{\sc \au{{Kempski}, Philipp}, \au{{Fielding}, Drummond~B.}, \au{{Quataert}, Eliot}, \au{{Galishnikova}, Alisa~K.}, \au{{Kunz}, Matthew~W.}, \au{{Philippov}, Alexander~A.} \& \au{{Ripperda}, Bart}} \yr{2023}  \at{{Cosmic ray transport in large-amplitude turbulence with small-scale field reversals}}.  \jt{\mnras}  \bvol{525}~(4),  \pg{4985--4998}.

\bibitem[{Kempski} {\em et~al.\/}(2025){Kempski}, {Li}, {Fielding}, {Quataert}, {Phinney}, {Kunz}, {Jow} \& {Philippov}]{Kempski2025a}
{\sc \au{{Kempski}, Philipp}, \au{{Li}, Dongzi}, \au{{Fielding}, Drummond~B.}, \au{{Quataert}, Eliot}, \au{{Phinney}, E.~Sterl}, \au{{Kunz}, Matthew~W.}, \au{{Jow}, Dylan~L.} \& \au{{Philippov}, Alexander~A.}} \yr{2025}  \at{{A Unified Model of Cosmic-Ray Propagation and Radio Extreme Scattering Events from Intermittent Interstellar Structures}}.  \jt{\apjl}  \bvol{990}~(1),  \pg{L18}.

\bibitem[{Kempski} \& {Quataert}(2022)]{Kempski2022}
{\sc \au{{Kempski}, Philipp} \& \au{{Quataert}, Eliot}} \yr{2022}  \at{{Reconciling cosmic ray transport theory with phenomenological models motivated by Milky-Way data}}.  \jt{\mnras}  \bvol{514}~(1),  \pg{657--674}.

\bibitem[{Khabarova} {\em et~al.\/}(2021){Khabarova}, {Malandraki}, {Malova}, {Kislov}, {Greco}, {Bruno}, {Pezzi}, {Servidio}, {Li}, {Matthaeus}, {Le Roux}, {Engelbrecht}, {Pecora}, {Zelenyi}, {Obridko} \& {Kuznetsov}]{Khabarova2021}
{\sc \au{{Khabarova}, O.}, \au{{Malandraki}, O.}, \au{{Malova}, H.}, \au{{Kislov}, R.}, \au{{Greco}, A.}, \au{{Bruno}, R.}, \au{{Pezzi}, O.}, \au{{Servidio}, S.}, \au{{Li}, Gang}, \au{{Matthaeus}, W.}, \au{{Le Roux}, J.}, \au{{Engelbrecht}, N.~E.}, \au{{Pecora}, F.}, \au{{Zelenyi}, L.}, \au{{Obridko}, V.} \& \au{{Kuznetsov}, V.}} \yr{2021}  \at{{Current Sheets, Plasmoids and Flux Ropes in the Heliosphere. Part I. 2-D or not 2-D? General and Observational Aspects}}.  \jt{\ssr}  \bvol{217}~(3),  \pg{38}.

\bibitem[{Krapivsky} {\em et~al.\/}(2010){Krapivsky}, {Redner} \& {Ben-Naim}]{Krapivsky2010}
{\sc \au{{Krapivsky}, Pavel~L.}, \au{{Redner}, Sidney} \& \au{{Ben-Naim}, Eli}} \yr{2010} {\em {A Kinetic View of Statistical Physics}\/}.  \publ{Cambridge, UK: Cambridge University Press}.

\bibitem[{Kulsrud}(2005)]{Kulsrud2005}
{\sc \au{{Kulsrud}, Russell~M.}} \yr{2005} {\em {Plasma Physics for Astrophysics}\/}.  \publ{Princeton, NJ: Princeton University Press}.

\bibitem[{Lemoine}(2021)]{Lemoine2021}
{\sc \au{{Lemoine}, Martin}} \yr{2021}  \at{{Particle acceleration in strong MHD turbulence}}.  \jt{\prd}  \bvol{104}~(6),  \pg{063020}.

\bibitem[{Lemoine}(2023)]{Lemoine2023}
{\sc \au{{Lemoine}, Martin}} \yr{2023}  \at{{Particle transport through localized interactions with sharp magnetic field bends in MHD turbulence}}.  \jt{Journal of Plasma Physics}  \bvol{89}~(5),  \pg{175890501}.

\bibitem[{Lesaffre} {\em et~al.\/}(2025){Lesaffre}, {Durrive}, {Goossaert}, {Poirier}, {Colombi}, {Richard}, {Allys} \& {Bethune}]{Lesaffre2025}
{\sc \au{{Lesaffre}, Pierre}, \au{{Durrive}, Jean-Baptiste}, \au{{Goossaert}, Jean}, \au{{Poirier}, Susie}, \au{{Colombi}, Stephane}, \au{{Richard}, Pablo}, \au{{Allys}, Erwan} \& \au{{Bethune}, William}} \yr{2025}  \at{{Multiscale Turbulence Synthesis: Validation in 2D Hydrodynamics}}.  \jt{arXiv e-prints}  \pg{p. arXiv:2506.23659}.

\bibitem[{Li} {\em et~al.\/}(2023){Li}, {Biferale}, {Bonaccorso}, {Scarpolini} \& {Buzzicotti}]{Li2023}
{\sc \au{{Li}, Tianyi}, \au{{Biferale}, Luca}, \au{{Bonaccorso}, Fabio}, \au{{Scarpolini}, Martino~Andrea} \& \au{{Buzzicotti}, Michele}} \yr{2023}  \at{{Synthetic Lagrangian Turbulence by Generative Diffusion Models}}.  \jt{arXiv e-prints}  \pg{p. arXiv:2307.08529}.

\bibitem[{Liang} \& {Oh}(2025)]{Liang2025}
{\sc \au{{Liang}, Naixin} \& \au{{Oh}, Siang~Peng}} \yr{2025}  \at{{L{\'e}vy Flights and Leaky Boxes: Anomalous Diffusion of Cosmic Rays}}.  \jt{arXiv e-prints}  \pg{p. arXiv:2503.10747}.

\bibitem[Lin {\em et~al.\/}(2021)Lin, Tian, Livescu \& Anghel]{lin-tian-etal:2021}
{\sc \au{Lin, Yen~Ting}, \au{Tian, Yifeng}, \au{Livescu, Daniel} \& \au{Anghel, Marian}} \yr{2021}  \at{Data-driven learning for the mori--zwanzig formalism: A generalization of the koopman learning framework}.  \jt{SIAM Journal on Applied Dynamical Systems}  \bvol{20}~(4),  \pg{2558--2601}.

\bibitem[{Loureiro} \& {Boldyrev}(2017)]{Loureiro2017}
{\sc \au{{Loureiro}, Nuno~F.} \& \au{{Boldyrev}, Stanislav}} \yr{2017}  \at{{Role of Magnetic Reconnection in Magnetohydrodynamic Turbulence}}.  \jt{\prl}  \bvol{118}~(24),  \pg{245101}.

\bibitem[{L{\"u}bke} {\em et~al.\/}(2024){L{\"u}bke}, {Effenberger}, {Wilbert}, {Fichtner} \& {Grauer}]{Lubke2024}
{\sc \au{{L{\"u}bke}, Jeremiah}, \au{{Effenberger}, Frederic}, \au{{Wilbert}, Mike}, \au{{Fichtner}, Horst} \& \au{{Grauer}, Rainer}} \yr{2024}  \at{{Towards synthetic magnetic turbulence with coherent structures}}.  \jt{EPL (Europhysics Letters)}  \bvol{146}~(4),  \pg{43001}.

\bibitem[{Maci} {\em et~al.\/}(2024){Maci}, {Keppens} \& {Bacchini}]{Maci2024}
{\sc \au{{Maci}, Daniela}, \au{{Keppens}, Rony} \& \au{{Bacchini}, Fabio}} \yr{2024}  \at{{BxC Toolkit: Generating Tailored Turbulent 3D Magnetic Fields}}.  \jt{\apjs}  \bvol{273}~(1),  \pg{11}.

\bibitem[{Magdziarz} \& {Weron}(2007)]{Magdziarz2007}
{\sc \au{{Magdziarz}, Marcin} \& \au{{Weron}, Aleksander}} \yr{2007}  \at{{Competition between subdiffusion and L{\'e}vy flights: A Monte Carlo approach}}.  \jt{\pre}  \bvol{75}~(5),  \pg{056702}.

\bibitem[{Malara} {\em et~al.\/}(2021){Malara}, {Perri} \& {Zimbardo}]{Malara2021}
{\sc \au{{Malara}, Francesco}, \au{{Perri}, Silvia} \& \au{{Zimbardo}, Gaetano}} \yr{2021}  \at{{Charged-particle chaotic dynamics in rotational discontinuities}}.  \jt{\pre}  \bvol{104}~(2),  \pg{025208}.

\bibitem[{Mallet} {\em et~al.\/}(2017){Mallet}, {Schekochihin} \& {Chandran}]{Mallet2017}
{\sc \au{{Mallet}, A.}, \au{{Schekochihin}, A.~A.} \& \au{{Chandran}, B.~D.~G.}} \yr{2017}  \at{{Disruption of sheet-like structures in Alfv{\'e}nic turbulence by magnetic reconnection}}.  \jt{\mnras}  \bvol{468}~(4),  \pg{4862--4871}.

\bibitem[{Martin} {\em et~al.\/}(2025){Martin}, {L{\"u}bke}, {Li}, {Buzzicotti}, {Grauer} \& {Biferale}]{Martin2025}
{\sc \au{{Martin}, Johannes}, \au{{L{\"u}bke}, Jeremiah}, \au{{Li}, Tianyi}, \au{{Buzzicotti}, Michele}, \au{{Grauer}, Rainer} \& \au{{Biferale}, Luca}} \yr{2025}  \at{{Generation of Cosmic-Ray Trajectories by a Diffusion Model Trained on Test Particles in 3D Magnetohydrodynamic Turbulence}}.  \jt{\apjs}  \bvol{277}~(2),  \pg{48}.

\bibitem[{Matthaeus} {\em et~al.\/}(2015){Matthaeus}, {Wan}, {Servidio}, {Greco}, {Osman}, {Oughton} \& {Dmitruk}]{Matthaeus2015}
{\sc \au{{Matthaeus}, W.~H.}, \au{{Wan}, M.}, \au{{Servidio}, S.}, \au{{Greco}, A.}, \au{{Osman}, K.~T.}, \au{{Oughton}, S.} \& \au{{Dmitruk}, P.}} \yr{2015}  \at{{Intermittency, nonlinear dynamics and dissipation in the solar wind and astrophysical plasmas}}.  \jt{Philosophical Transactions of the Royal Society of London Series A}  \bvol{373}~(2041),  \pg{20140154--20140154}.

\bibitem[{Mertsch}(2020)]{Mertsch2020}
{\sc \au{{Mertsch}, Philipp}} \yr{2020}  \at{{Test particle simulations of cosmic rays}}.  \jt{\apss}  \bvol{365}~(8),  \pg{135}.

\bibitem[{Metzler} \& {Klafter}(2000)]{Metzler2000}
{\sc \au{{Metzler}, Ralf} \& \au{{Klafter}, Joseph}} \yr{2000}  \at{{The random walk's guide to anomalous diffusion: a fractional dynamics approach}}.  \jt{\physrep}  \bvol{339}~(1),  \pg{1--77}.

\bibitem[{Meyrand} {\em et~al.\/}(2016){Meyrand}, {Galtier} \& {Kiyani}]{Meyrand2016}
{\sc \au{{Meyrand}, Romain}, \au{{Galtier}, S{\'e}bastien} \& \au{{Kiyani}, Khurom~H.}} \yr{2016}  \at{{Direct Evidence of the Transition from Weak to Strong Magnetohydrodynamic Turbulence}}.  \jt{\prl}  \bvol{116}~(10),  \pg{105002}.

\bibitem[Miller(2019)]{MillerPhd2019}
{\sc \au{Miller, Daniel}} \yr{2019}  \at{Alignment and structure in mhd dynamos}. PhD thesis, University of Exeter.

\bibitem[Mininni {\em et~al.\/}(2006)Mininni, Pouquet \& Montgomery]{minini-pouquet-etal:2006}
{\sc \au{Mininni, P.~D.}, \au{Pouquet, A.~G.} \& \au{Montgomery, D.~C.}} \yr{2006}  \at{Small-scale structures in three-dimensional magnetohydrodynamic turbulence}.  \jt{Phys. Rev. Lett.}  \bvol{97},  \pg{244503}.

\bibitem[{Minnie} {\em et~al.\/}(2009){Minnie}, {Matthaeus}, {Bieber}, {Ruffolo} \& {Burger}]{Minnie2009}
{\sc \au{{Minnie}, J.}, \au{{Matthaeus}, W.~H.}, \au{{Bieber}, J.~W.}, \au{{Ruffolo}, D.} \& \au{{Burger}, R.~A.}} \yr{2009}  \at{{When do particles follow field lines?}}  \jt{Journal of Geophysical Research (Space Physics)}  \bvol{114}~(A1),  \pg{A01102}.

\bibitem[{Morillo} \& {Alexakis}(2025)]{Morillo2025}
{\sc \au{{Morillo}, Jos{\'e} Mar{\'\i}a~Garc{\'\i}a} \& \au{{Alexakis}, Alexandros}} \yr{2025}  \at{{Magnetic reconnection, plasmoids and numerical resolution}}.  \jt{Journal of Fluid Mechanics}  \bvol{1007},  \pg{R3}.

\bibitem[{Neuer} \& {Spatschek}(2006)]{Neuer2006}
{\sc \au{{Neuer}, Marcus} \& \au{{Spatschek}, Karl~H.}} \yr{2006}  \at{{Diffusion of test particles in stochastic magnetic fields for small Kubo numbers}}.  \jt{\pre}  \bvol{73}~(2),  \pg{026404}.

\bibitem[{Ntormousi} {\em et~al.\/}(2024){Ntormousi}, {Vlahos}, {Konstantinou} \& {Isliker}]{Ntormousi2024}
{\sc \au{{Ntormousi}, Evangelia}, \au{{Vlahos}, Loukas}, \au{{Konstantinou}, Anna} \& \au{{Isliker}, Heinz}} \yr{2024}  \at{{Strong turbulence and magnetic coherent structures in the interstellar medium}}.  \jt{\aap}  \bvol{691},  \pg{A149}.

\bibitem[{Parker}(1965)]{Parker1965}
{\sc \au{{Parker}, E.~N.}} \yr{1965}  \at{{The passage of energetic charged particles through interplanetary space}}.  \jt{\planss}  \bvol{13}~(1),  \pg{9--49}.

\bibitem[{Perez} \& {Boldyrev}(2009)]{Perez2009}
{\sc \au{{Perez}, Jean~Carlos} \& \au{{Boldyrev}, Stanislav}} \yr{2009}  \at{{Role of Cross-Helicity in Magnetohydrodynamic Turbulence}}.  \jt{\prl}  \bvol{102}~(2),  \pg{025003}.

\bibitem[{Pezzi} \& {Blasi}(2024)]{Pezzi2024}
{\sc \au{{Pezzi}, O.} \& \au{{Blasi}, P.}} \yr{2024}  \at{{Galactic cosmic ray transport in the absence of resonant scattering}}.  \jt{\mnras}  \bvol{529}~(1),  \pg{L13--L18}.

\bibitem[{Pezzi} {\em et~al.\/}(2022){Pezzi}, {Blasi} \& {Matthaeus}]{Pezzi2022}
{\sc \au{{Pezzi}, Oreste}, \au{{Blasi}, Pasquale} \& \au{{Matthaeus}, William~H.}} \yr{2022}  \at{{Relativistic Particle Transport and Acceleration in Structured Plasma Turbulence}}.  \jt{\apj}  \bvol{928}~(1),  \pg{25}.

\bibitem[{Pfrommer} {\em et~al.\/}(2017){Pfrommer}, {Pakmor}, {Schaal}, {Simpson} \& {Springel}]{Pfrommer2017}
{\sc \au{{Pfrommer}, C.}, \au{{Pakmor}, R.}, \au{{Schaal}, K.}, \au{{Simpson}, C.~M.} \& \au{{Springel}, V.}} \yr{2017}  \at{{Simulating cosmic ray physics on a moving mesh}}.  \jt{\mnras}  \bvol{465}~(4),  \pg{4500--4529}.

\bibitem[{Politano} \& {Pouquet}(1995)]{Politano1995b}
{\sc \au{{Politano}, H.} \& \au{{Pouquet}, A.}} \yr{1995}  \at{{Model of intermittency in magnetohydrodynamic turbulence}}.  \jt{\pre}  \bvol{52}~(1),  \pg{636--641}.

\bibitem[{Pucci} {\em et~al.\/}(2016){Pucci}, {Malara}, {Perri}, {Zimbardo}, {Sorriso-Valvo} \& {Valentini}]{Pucci2016}
{\sc \au{{Pucci}, F.}, \au{{Malara}, F.}, \au{{Perri}, S.}, \au{{Zimbardo}, G.}, \au{{Sorriso-Valvo}, L.} \& \au{{Valentini}, F.}} \yr{2016}  \at{{Energetic particle transport in the presence of magnetic turbulence: influence of spectral extension and intermittency}}.  \jt{\mnras}  \bvol{459}~(3),  \pg{3395--3406}.

\bibitem[{Pugliese} {\em et~al.\/}(2023){Pugliese}, {Brodiano}, {Andr{\'e}s} \& {Dmitruk}]{Pugliese2023}
{\sc \au{{Pugliese}, F.}, \au{{Brodiano}, M.}, \au{{Andr{\'e}s}, N.} \& \au{{Dmitruk}, P.}} \yr{2023}  \at{{Energization of Charged Test Particles in Magnetohydrodynamic Fields: Waves versus Turbulence Picture}}.  \jt{\apj}  \bvol{959}~(1),  \pg{28}.

\bibitem[{Qin} {\em et~al.\/}(2002){Qin}, {Matthaeus} \& {Bieber}]{Qin2002a}
{\sc \au{{Qin}, G.}, \au{{Matthaeus}, W.~H.} \& \au{{Bieber}, J.~W.}} \yr{2002}  \at{{Perpendicular Transport of Charged Particles in Composite Model Turbulence: Recovery of Diffusion}}.  \jt{\apjl}  \bvol{578}~(2),  \pg{L117--L120}.

\bibitem[{Reichherzer} {\em et~al.\/}(2020){Reichherzer}, {Becker Tjus}, {Zweibel}, {Merten} \& {Pueschel}]{Reichherzer2020}
{\sc \au{{Reichherzer}, P.}, \au{{Becker Tjus}, J.}, \au{{Zweibel}, E.~G.}, \au{{Merten}, L.} \& \au{{Pueschel}, M.~J.}} \yr{2020}  \at{{Turbulence-level dependence of cosmic ray parallel diffusion}}.  \jt{\mnras}  \bvol{498}~(4),  \pg{5051--5064}.

\bibitem[{Reichherzer} {\em et~al.\/}(2025){Reichherzer}, {Bott}, {Ewart}, {Gregori}, {Kempski}, {Kunz} \& {Schekochihin}]{Reichherzer2025}
{\sc \au{{Reichherzer}, Patrick}, \au{{Bott}, Archie F.~A.}, \au{{Ewart}, Robert~J.}, \au{{Gregori}, Gianluca}, \au{{Kempski}, Philipp}, \au{{Kunz}, Matthew~W.} \& \au{{Schekochihin}, Alexander~A.}} \yr{2025}  \at{{Efficient micromirror confinement of sub-teraelectronvolt cosmic rays in galaxy clusters}}.  \jt{Nature Astronomy}  \bvol{9},  \pg{438--448}.

\bibitem[{Rieder} \& {Teyssier}(2017)]{Rieder2017}
{\sc \au{{Rieder}, Michael} \& \au{{Teyssier}, Romain}} \yr{2017}  \at{{A small-scale dynamo in feedback-dominated galaxies - II. The saturation phase and the final magnetic configuration}}.  \jt{\mnras}  \bvol{471}~(3),  \pg{2674--2686}.

\bibitem[{Rincon}(2019)]{Rincon2019}
{\sc \au{{Rincon}, Fran{\c{c}}ois}} \yr{2019}  \at{{Dynamo theories}}.  \jt{Journal of Plasma Physics}  \bvol{85}~(4),  \pg{205850401}.

\bibitem[{Ripperda} {\em et~al.\/}(2018){Ripperda}, {Bacchini}, {Teunissen}, {Xia}, {Porth}, {Sironi}, {Lapenta} \& {Keppens}]{Ripperda2018}
{\sc \au{{Ripperda}, B.}, \au{{Bacchini}, F.}, \au{{Teunissen}, J.}, \au{{Xia}, C.}, \au{{Porth}, O.}, \au{{Sironi}, L.}, \au{{Lapenta}, G.} \& \au{{Keppens}, R.}} \yr{2018}  \at{{A Comprehensive Comparison of Relativistic Particle Integrators}}.  \jt{\apjs}  \bvol{235}~(1),  \pg{21}.

\bibitem[{Robitaille} {\em et~al.\/}(2020){Robitaille}, {Abdeldayem}, {Joncour}, {Moraux}, {Motte}, {Lesaffre} \& {Khalil}]{Robitaille2020}
{\sc \au{{Robitaille}, J.~F.}, \au{{Abdeldayem}, A.}, \au{{Joncour}, I.}, \au{{Moraux}, E.}, \au{{Motte}, F.}, \au{{Lesaffre}, P.} \& \au{{Khalil}, A.}} \yr{2020}  \at{{Statistical model for filamentary structures of molecular clouds. The modified multiplicative random cascade model and its multifractal nature}}.  \jt{\aap}  \bvol{641},  \pg{A138}.

\bibitem[{Ryu} {\em et~al.\/}(2012){Ryu}, {Schleicher}, {Treumann}, {Tsagas} \& {Widrow}]{Ryu2012}
{\sc \au{{Ryu}, D.}, \au{{Schleicher}, D.~R.~G.}, \au{{Treumann}, R.~A.}, \au{{Tsagas}, C.~G.} \& \au{{Widrow}, L.~M.}} \yr{2012}  \at{{Magnetic Fields in the Large-Scale Structure of the Universe}}.  \jt{\ssr}  \bvol{166}~(1-4),  \pg{1--35}.

\bibitem[{Sampson} {\em et~al.\/}(2023){Sampson}, {Beattie}, {Krumholz}, {Crocker}, {Federrath} \& {Seta}]{Sampson2023}
{\sc \au{{Sampson}, Matt~L.}, \au{{Beattie}, James~R.}, \au{{Krumholz}, Mark~R.}, \au{{Crocker}, Roland~M.}, \au{{Federrath}, Christoph} \& \au{{Seta}, Amit}} \yr{2023}  \at{{Turbulent diffusion of streaming cosmic rays in compressible, partially ionized plasma}}.  \jt{\mnras}  \bvol{519}~(1),  \pg{1503--1525}.

\bibitem[{Schekochihin}(2022)]{Schekochihin2022}
{\sc \au{{Schekochihin}, Alexander~A.}} \yr{2022}  \at{{MHD turbulence: a biased review}}.  \jt{Journal of Plasma Physics}  \bvol{88}~(5),  \pg{155880501}.

\bibitem[{Schekochihin} {\em et~al.\/}(2004){Schekochihin}, {Cowley}, {Taylor}, {Maron} \& {McWilliams}]{Schekochihin2004a}
{\sc \au{{Schekochihin}, Alexander~A.}, \au{{Cowley}, Steven~C.}, \au{{Taylor}, Samuel~F.}, \au{{Maron}, Jason~L.} \& \au{{McWilliams}, James~C.}} \yr{2004}  \at{{Simulations of the Small-Scale Turbulent Dynamo}}.  \jt{\apj}  \bvol{612}~(1),  \pg{276--307}.

\bibitem[{Schekochihin} {\em et~al.\/}(2002){Schekochihin}, {Maron}, {Cowley} \& {McWilliams}]{Schekochihin2002b}
{\sc \au{{Schekochihin}, Alexander~A.}, \au{{Maron}, Jason~L.}, \au{{Cowley}, Steven~C.} \& \au{{McWilliams}, James~C.}} \yr{2002}  \at{{The Small-Scale Structure of Magnetohydrodynamic Turbulence with Large Magnetic Prandtl Numbers}}.  \jt{\apj}  \bvol{576}~(2),  \pg{806--813}.

\bibitem[{Schlickeiser}(2002)]{Schlickeiser2002}
{\sc \au{{Schlickeiser}, Reinhard}} \yr{2002} {\em {Cosmic Ray Astrophysics}\/}.  \publ{Berlin, Germany: Springer}.

\bibitem[Schorlepp \& Grafke(2025)]{schorlepp-grafke:2025}
{\sc \au{Schorlepp, Timo} \& \au{Grafke, Tobias}} \yr{2025}  \at{Scalability of the second-order reliability method for stochastic differential equations with multiplicative noise}.  \jt{arXiv preprint arXiv:2502.20114}.

\bibitem[{Servidio} {\em et~al.\/}(2011){Servidio}, {Dmitruk}, {Greco}, {Wan}, {Donato}, {Cassak}, {Shay}, {Carbone} \& {Matthaeus}]{Servidio2011}
{\sc \au{{Servidio}, S.}, \au{{Dmitruk}, P.}, \au{{Greco}, A.}, \au{{Wan}, M.}, \au{{Donato}, S.}, \au{{Cassak}, P.~A.}, \au{{Shay}, M.~A.}, \au{{Carbone}, V.} \& \au{{Matthaeus}, W.~H.}} \yr{2011}  \at{{Magnetic reconnection as an element of turbulence}}.  \jt{Nonlinear Processes in Geophysics}  \bvol{18}~(5),  \pg{675--695}.

\bibitem[{Seta} {\em et~al.\/}(2020){Seta}, {Bushby}, {Shukurov} \& {Wood}]{Seta2020}
{\sc \au{{Seta}, Amit}, \au{{Bushby}, Paul~J.}, \au{{Shukurov}, Anvar} \& \au{{Wood}, Toby~S.}} \yr{2020}  \at{{Saturation mechanism of the fluctuation dynamo at Pr$_{M}$ {\ensuremath{\geq}} 1}}.  \jt{Physical Review Fluids}  \bvol{5}~(4),  \pg{043702}.

\bibitem[{Shukurov} {\em et~al.\/}(2017){Shukurov}, {Snodin}, {Seta}, {Bushby} \& {Wood}]{Shukurov2017}
{\sc \au{{Shukurov}, Anvar}, \au{{Snodin}, Andrew~P.}, \au{{Seta}, Amit}, \au{{Bushby}, Paul~J.} \& \au{{Wood}, Toby~S.}} \yr{2017}  \at{{Cosmic Rays in Intermittent Magnetic Fields}}.  \jt{\apjl}  \bvol{839}~(1),  \pg{L16}.

\bibitem[{Sironi} {\em et~al.\/}(2023){Sironi}, {Comisso} \& {Golant}]{Sironi2023}
{\sc \au{{Sironi}, Lorenzo}, \au{{Comisso}, Luca} \& \au{{Golant}, Ryan}} \yr{2023}  \at{{Generation of Near-Equipartition Magnetic Fields in Turbulent Collisionless Plasmas}}.  \jt{\prl}  \bvol{131}~(5),  \pg{055201}.

\bibitem[{Skilling}(1975)]{Skilling1975c}
{\sc \au{{Skilling}, J.}} \yr{1975}  \at{{Cosmic ray streaming - I. Effect of Alfv{\'e}n waves on particles.}}  \jt{\mnras}  \bvol{172},  \pg{557--566}.

\bibitem[{St-Onge} \& {Kunz}(2018)]{St-Onge2018}
{\sc \au{{St-Onge}, Denis~A.} \& \au{{Kunz}, Matthew~W.}} \yr{2018}  \at{{Fluctuation Dynamo in a Collisionless, Weakly Magnetized Plasma}}.  \jt{\apjl}  \bvol{863}~(2),  \pg{L25}.

\bibitem[{Steinwandel} {\em et~al.\/}(2024){Steinwandel}, {Dolag}, {B{\"o}ss} \& {Marin-Gilabert}]{Steinwandel2024}
{\sc \au{{Steinwandel}, Ulrich~P.}, \au{{Dolag}, Klaus}, \au{{B{\"o}ss}, Ludwig~M.} \& \au{{Marin-Gilabert}, Tirso}} \yr{2024}  \at{{Toward Cosmological Simulations of the Magnetized Intracluster Medium with Resolved Coulomb Collision Scale}}.  \jt{\apj}  \bvol{967}~(2),  \pg{125}.

\bibitem[{Strauss} \& {Effenberger}(2017)]{Strauss2017}
{\sc \au{{Strauss}, R. Du~Toit} \& \au{{Effenberger}, Frederic}} \yr{2017}  \at{{A Hitch-hiker's Guide to Stochastic Differential Equations. Solution Methods for Energetic Particle Transport in Space Physics and Astrophysics}}.  \jt{\ssr}  \bvol{212}~(1-2),  \pg{151--192}.

\bibitem[{Subedi} {\em et~al.\/}(2014){Subedi}, {Chhiber}, {Tessein}, {Wan} \& {Matthaeus}]{Subedi2014}
{\sc \au{{Subedi}, P.}, \au{{Chhiber}, R.}, \au{{Tessein}, J.~A.}, \au{{Wan}, M.} \& \au{{Matthaeus}, W.~H.}} \yr{2014}  \at{{Generating Synthetic Magnetic Field Intermittency Using a Minimal Multiscale Lagrangian Mapping Approach}}.  \jt{\apj}  \bvol{796}~(2),  \pg{97}.

\bibitem[{Tu} \& {Marsch}(1995)]{tu-marsch:1995}
{\sc \au{{Tu}, C.~Y.} \& \au{{Marsch}, E.}} \yr{1995}  \at{{Magnetohydrodynamic Structures Waves and Turbulence in the Solar Wind - Observations and Theories}}.  \jt{\ssr}  \bvol{73}~(1-2),  \pg{1--210}.

\bibitem[{van den Berg} {\em et~al.\/}(2024){van den Berg}, {Els} \& {Engelbrecht}]{vandenBerg2024}
{\sc \au{{van den Berg}, J.~P.}, \au{{Els}, P.~L.} \& \au{{Engelbrecht}, N.~E.}} \yr{2024}  \at{{An Evaluation of Different Numerical Methods to Calculate the Pitch-angle Diffusion Coefficient from Full-orbit Simulations: Disentangling a Rope of Sand}}.  \jt{\apj}  \bvol{977}~(2),  \pg{174}.

\bibitem[{Vazza} {\em et~al.\/}(2018){Vazza}, {Brunetti}, {Br{\"u}ggen} \& {Bonafede}]{Vazza2018}
{\sc \au{{Vazza}, F.}, \au{{Brunetti}, G.}, \au{{Br{\"u}ggen}, M.} \& \au{{Bonafede}, A.}} \yr{2018}  \at{{Resolved magnetic dynamo action in the simulated intracluster medium}}.  \jt{\mnras}  \bvol{474}~(2),  \pg{1672--1687}.

\bibitem[{Vinogradov} {\em et~al.\/}(2024){Vinogradov}, {Alexandrova}, {D{\'e}moulin}, {Artemyev}, {Maksimovic}, {Mangeney}, {Vasiliev}, {Petrukovich} \& {Bale}]{Vinogradov2024}
{\sc \au{{Vinogradov}, Alexander}, \au{{Alexandrova}, Olga}, \au{{D{\'e}moulin}, Pascal}, \au{{Artemyev}, Anton}, \au{{Maksimovic}, Milan}, \au{{Mangeney}, Andr{\'e}}, \au{{Vasiliev}, Alexei}, \au{{Petrukovich}, Anatoli~A.} \& \au{{Bale}, Stuart}} \yr{2024}  \at{{Embedded Coherent Structures from Magnetohydrodynamics to Sub-ion Scales in Turbulent Solar Wind at 0.17 au}}.  \jt{\apj}  \bvol{971}~(1),  \pg{88}.

\bibitem[{Wang} {\em et~al.\/}(2024){Wang}, {Chhiber}, {Roy}, {Cuesta}, {Pecora}, {Yang}, {Fu}, {Li} \& {Matthaeus}]{Wang2024}
{\sc \au{{Wang}, Jiaming}, \au{{Chhiber}, Rohit}, \au{{Roy}, Sohom}, \au{{Cuesta}, Manuel~E.}, \au{{Pecora}, Francesco}, \au{{Yang}, Yan}, \au{{Fu}, Xiangrong}, \au{{Li}, Hui} \& \au{{Matthaeus}, William~H.}} \yr{2024}  \at{{Anisotropy of Density Fluctuations in the Solar Wind at 1 au}}.  \jt{\apj}  \bvol{967}~(2),  \pg{150}.

\bibitem[{Weygand} {\em et~al.\/}(2011){Weygand}, {Matthaeus}, {Dasso} \& {Kivelson}]{Weygand-etal-2011}
{\sc \au{{Weygand}, James~M.}, \au{{Matthaeus}, W.~H.}, \au{{Dasso}, S.} \& \au{{Kivelson}, M.~G.}} \yr{2011}  \at{{Correlation and Taylor scale variability in the interplanetary magnetic field fluctuations as a function of solar wind speed}}.  \jt{Journal of Geophysical Research (Space Physics)}  \bvol{116}~(A8),  \pg{A08102}.

\bibitem[{Weygand} {\em et~al.\/}(2013){Weygand}, {Matthaeus}, {Kivelson} \& {Dasso}]{Weygand-etal-2013}
{\sc \au{{Weygand}, James~M.}, \au{{Matthaeus}, W.~H.}, \au{{Kivelson}, M.~G.} \& \au{{Dasso}, S.}} \yr{2013}  \at{{Magnetic correlation functions in the slow and fast solar wind in the Eulerian reference frame}}.  \jt{Journal of Geophysical Research (Space Physics)}  \bvol{118}~(7),  \pg{3995--4004}.

\bibitem[Wilbert(2023)]{WilbertPhd2023}
{\sc \au{Wilbert, Mike}} \yr{2023}  \at{Implementation and application of a pseudo-spectral mhd solver combined with an immersed boundary method to support the dresdyn dynamo experiment}. PhD thesis, Ruhr-Universität Bochum.

\bibitem[{Wilbert} {\em et~al.\/}(2022){Wilbert}, {Giesecke} \& {Grauer}]{Wilbert2022}
{\sc \au{{Wilbert}, Mike}, \au{{Giesecke}, Andr{\'e}} \& \au{{Grauer}, Rainer}} \yr{2022}  \at{{Numerical investigation of the flow inside a precession-driven cylindrical cavity with additional baffles using an immersed boundary method}}.  \jt{Physics of Fluids}  \bvol{34}~(9),  \pg{096607}.

\bibitem[{Xu} \& {Yan}(2013)]{Xu2013}
{\sc \au{{Xu}, Siyao} \& \au{{Yan}, Huirong}} \yr{2013}  \at{{Cosmic-Ray Parallel and Perpendicular Transport in Turbulent Magnetic Fields}}.  \jt{\apj}  \bvol{779}~(2),  \pg{140}.

\bibitem[{Yang} {\em et~al.\/}(2019){Yang}, {Wan}, {Matthaeus}, {Shi}, {Parashar}, {Lu} \& {Chen}]{Yang2019}
{\sc \au{{Yang}, Yan}, \au{{Wan}, Minping}, \au{{Matthaeus}, William~H.}, \au{{Shi}, Yipeng}, \au{{Parashar}, Tulasi~N.}, \au{{Lu}, Quanming} \& \au{{Chen}, Shiyi}} \yr{2019}  \at{{Role of magnetic field curvature in magnetohydrodynamic turbulence}}.  \jt{Physics of Plasmas}  \bvol{26}~(7),  \pg{072306}.

\bibitem[{Zaburdaev} {\em et~al.\/}(2015){Zaburdaev}, {Denisov} \& {Klafter}]{Zaburdaev2015}
{\sc \au{{Zaburdaev}, V.}, \au{{Denisov}, S.} \& \au{{Klafter}, J.}} \yr{2015}  \at{{L{\'e}vy walks}}.  \jt{Reviews of Modern Physics}  \bvol{87}~(2),  \pg{483--530}.

\bibitem[{Zhang} \& {Xu}(2024)]{Zhang2024}
{\sc \au{{Zhang}, Chao} \& \au{{Xu}, Siyao}} \yr{2024}  \at{{Cosmic Ray Diffusion in Magnetic Fields Amplified by Nonlinear Turbulent Dynamo}}.  \jt{\apj}  \bvol{975}~(1),  \pg{65}.

\bibitem[{Zhdankin} {\em et~al.\/}(2016){Zhdankin}, {Boldyrev} \& {Uzdensky}]{Zhdankin2016}
{\sc \au{{Zhdankin}, Vladimir}, \au{{Boldyrev}, Stanislav} \& \au{{Uzdensky}, Dmitri~A.}} \yr{2016}  \at{{Scalings of intermittent structures in magnetohydrodynamic turbulence}}.  \jt{Physics of Plasmas}  \bvol{23}~(5),  \pg{055705}.

\bibitem[{Zhou} {\em et~al.\/}(2024){Zhou}, {Zhdankin}, {Kunz}, {Loureiro} \& {Uzdensky}]{Zhou2024}
{\sc \au{{Zhou}, Muni}, \au{{Zhdankin}, Vladimir}, \au{{Kunz}, Matthew~W.}, \au{{Loureiro}, Nuno~F.} \& \au{{Uzdensky}, Dmitri~A.}} \yr{2024}  \at{{Magnetogenesis in a Collisionless Plasma: From Weibel Instability to Turbulent Dynamo}}.  \jt{\apj}  \bvol{960}~(1),  \pg{12}.

\bibitem[{Zimbardo} \& {Perri}(2020)]{Zimbardo2020}
{\sc \au{{Zimbardo}, Gaetano} \& \au{{Perri}, Silvia}} \yr{2020}  \at{{Non-Markovian Pitch-angle Scattering as the Origin of Particle Superdiffusion Parallel to the Magnetic Field}}.  \jt{\apj}  \bvol{903}~(2),  \pg{105}.

\end{thebibliography}

\end{document}